%% file: output.tex
\documentclass[usenatbib]{mnras}

\pdfoutput=1 
\usepackage{newtxtext,newtxmath}
\usepackage{threeparttable}
\usepackage{tikz}
\usepackage{longtable}

\usepackage[T1]{fontenc}

\hypersetup{linkcolor=red}          

\newcommand{\HL}[1]{\textcolor{black}{#1}}

\def\equationautorefname~#1\null{equation~(#1)\null}

\usepackage{scalerel,tikz}
\usetikzlibrary{svg.path}
\definecolor{orcidlogocol}{HTML}{A6CE39}
\tikzset{orcidlogo/.pic={
 \fill[orcidlogocol] svg{M256,128c0,70.7-57.3,128-128,128C57.3,256,0,198.7,0,128C0,57.3,57.3,0,128,0C198.7,0,256,57.3,256,128z};
 \fill[white] svg{M86.3,186.2H70.9V79.1h15.4v48.4V186.2z}
 svg{M108.9,79.1h41.6c39.6,0,57,28.3,57,53.6c0,27.5-21.5,53.6-56.8,53.6h-41.8V79.1z M124.3,172.4h24.5c34.9,0,42.9-26.5,42.9-39.7c0-21.5-13.7-39.7-43.7-39.7h-23.7V172.4z}
 svg{M88.7,56.8c0,5.5-4.5,10.1-10.1,10.1c-5.6,0-10.1-4.6-10.1-10.1c0-5.6,4.5-10.1,10.1-10.1C84.2,46.7,88.7,51.3,88.7,56.8z};
}}
\newcommand\orcidicon[1]{\href{https://orcid.org/#1}{\mbox{\scalerel*{
\begin{tikzpicture}[yscale=-1,transform shape]
\pic{orcidlogo};
\end{tikzpicture}
}{|}}}}

\title[A Catalogue of broad-line MaNGA galaxies]{A complete catalogue of broad-line AGNs and double-peaked emission lines from MaNGA integral-field spectroscopy of 10K galaxies: stellar population of AGNs, supermassive black holes, and dual AGNs}

\author[Youquan Fu. et al ]{
Youquan Fu\orcidicon{0009-0003-3720-6870},$^{1}$\thanks{E-mail: fuyq22@mails.tsinghua.edu.cn}
Michele Cappellari\orcidicon{0000-0002-1283-8420},$^{2}$
Shude Mao\orcidicon{0000-0001-8317-2788},$^{1,3}$
Shengdong Lu\orcidicon{0000-0002-6726-9499},$^{1}$
Kai Zhu\orcidicon{0000-0002-2583-2669}, $^{3,4,5}$
Ran Li\orcidicon{0000-0003-3899-0612}$^{3,4,5}$
\\
$^{1}$Department of Astronomy, Tsinghua University, Beijing 100084, China\\
$^{2}$Sub-Department of Astrophysics, Department of Physics, University of Oxford, Denys Wilkinson Building, Keble Road, Oxford, OX1 3RH, UK\\
$^{3}$National Astronomical Observatories, Chinese Academy of Sciences, 20A Datun Road, Chaoyang District, Beijing 100101, China\\
$^{4}$Institute for Frontiers in Astronomy and Astrophysics, Beijing Normal University, Beijing 102206, China\\
$^{5}$School of Astronomy and Space Science, University of Chinese Academy of Sciences, Beijing 100049, China\\
}

\date{Accepted 2023 July 17. Received 2023 July 12; in original form 2023 April 25}

\pubyear{2023}

\begin{document}
\label{firstpage}
\pagerange{\pageref{firstpage}--\pageref{lastpage}}
\maketitle
\begin{abstract}
We analyse the integral-field spectroscopy data for the $\approx10,000$ galaxies in final data release of the MaNGA survey. We identify 188 galaxies for which the emission lines cannot be described by single Gaussian components. These galaxies can be classified into (1) 38 galaxies with broad $\rm H\alpha$ and [OIII] $\rm \lambda$5007 lines, (2) 101 galaxies with broad $\rm H\alpha$ lines but no broad [OIII] $\rm \lambda$5007 lines, and (3) 49 galaxies with double-peaked narrow emission lines. Most of the broad line galaxies are classified as Active Galactic Nuclei (AGN) from their line ratios. The catalogue helps us further understand the AGN-galaxy coevolution through the stellar population of broad-line region host galaxies and the relation between broad lines' properties and the host galaxies' dynamical properties. The stellar population properties (including mass, age and metallicity) of broad-line host galaxies suggest there is no significant difference between narrow-line Seyfert-2 galaxies and Type-1 AGN with broad $\rm H\alpha$ lines.  We use the broad-$\rm H\alpha$ line width and luminosity to estimate masses of black hole in these galaxies, and test the $M_{\rm BH}-\sigma_{\rm e}$ relation in Type-1 AGN host galaxies. Furthermore we find three dual AGN candidates supported by radio images from the VLA FIRST survey. This sample may be useful for further studies on AGN activities and feedback processes.
\end{abstract}




\section{Introduction}
\input{introduction}
\section{Sample Selection}
\label{sec:sample}
\input{Sample}

\section{Methods}
\label{sec:methods}
\begin{figure*}
	\includegraphics[width=\textwidth]{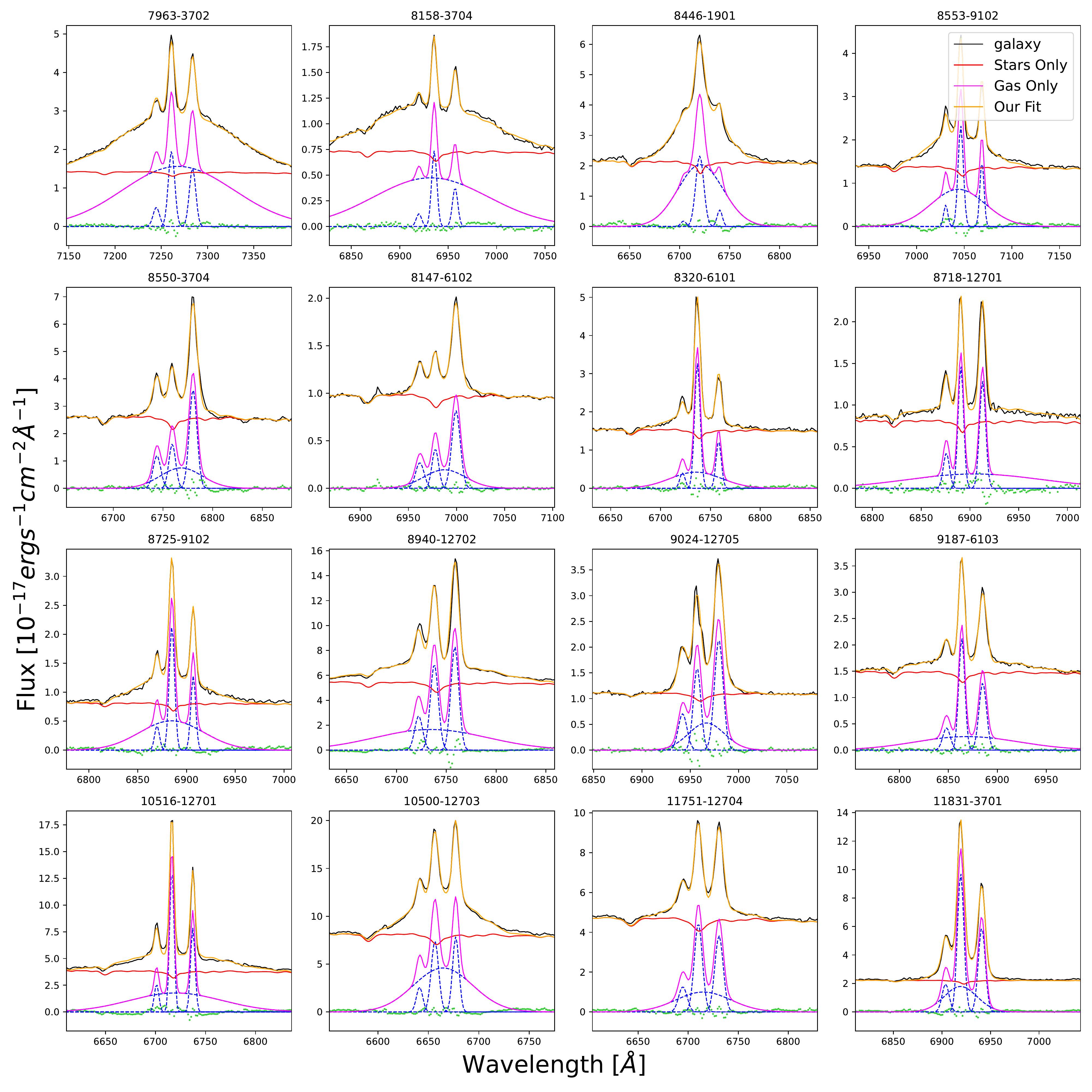}
    \caption{Examples of galaxies with broad $\rm H\alpha$ lines. The black spectra are the observed galaxy flux in $\rm H\alpha$ region. The orange spectra are our multi-Gaussian fit. In the lower half of the picture, the green dots represent the residual of multi-Gaussian fit. Individual Gaussian line fitting is shown in dashed blue lines. The total flux of gas and stellar are shown in magenta and red lines.}
    \label{fig:egbroadha}
\end{figure*}

\begin{figure*}
	\includegraphics[width=\textwidth]{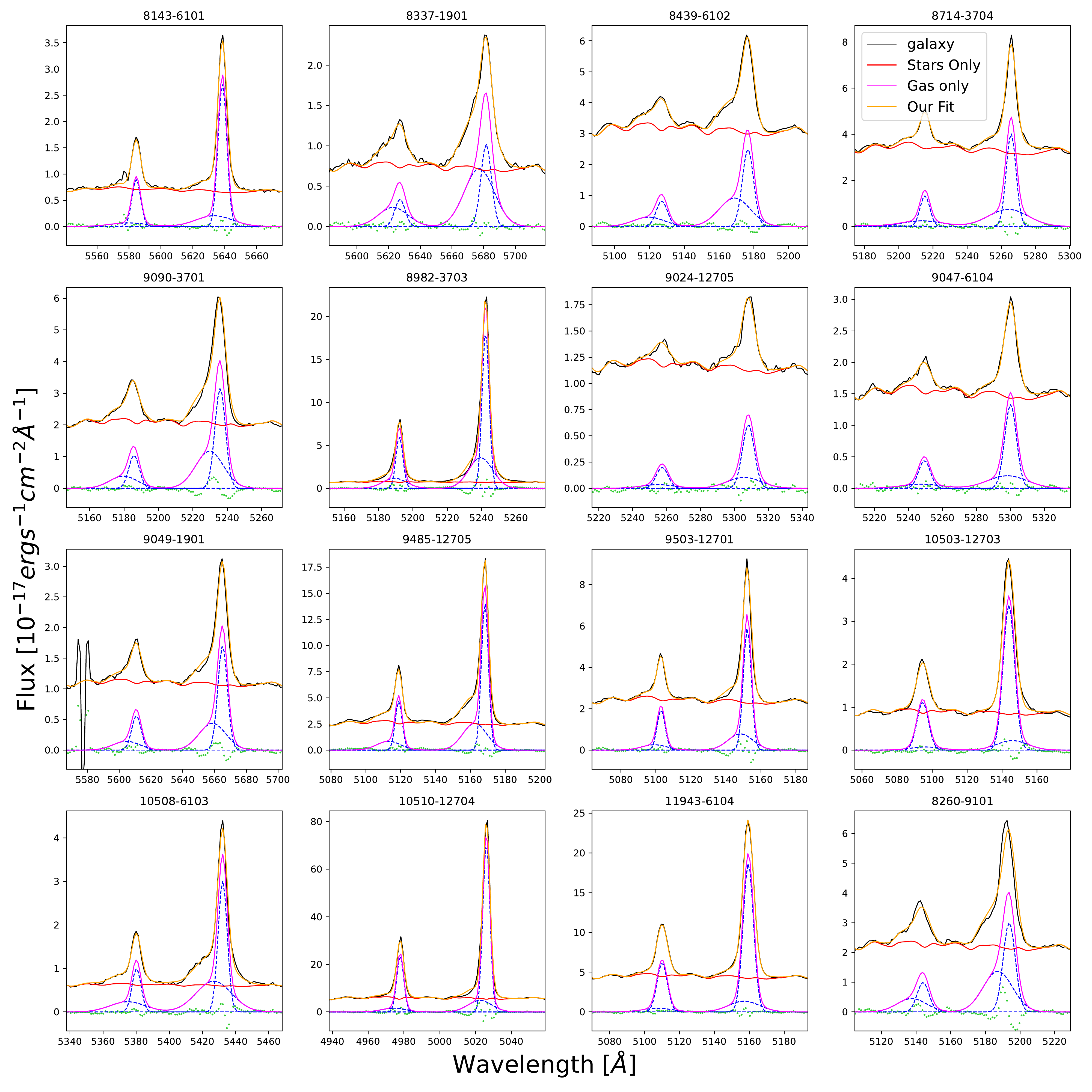}
    \caption{Examples of galaxies with broad $\rm [OIII]$ doublets. Correspondence between spectrum and colour is described in \autoref{fig:egbroadha}. }
    \label{fig:egbroadOIII}
\end{figure*}

\begin{figure*}
	\includegraphics[width=\textwidth]{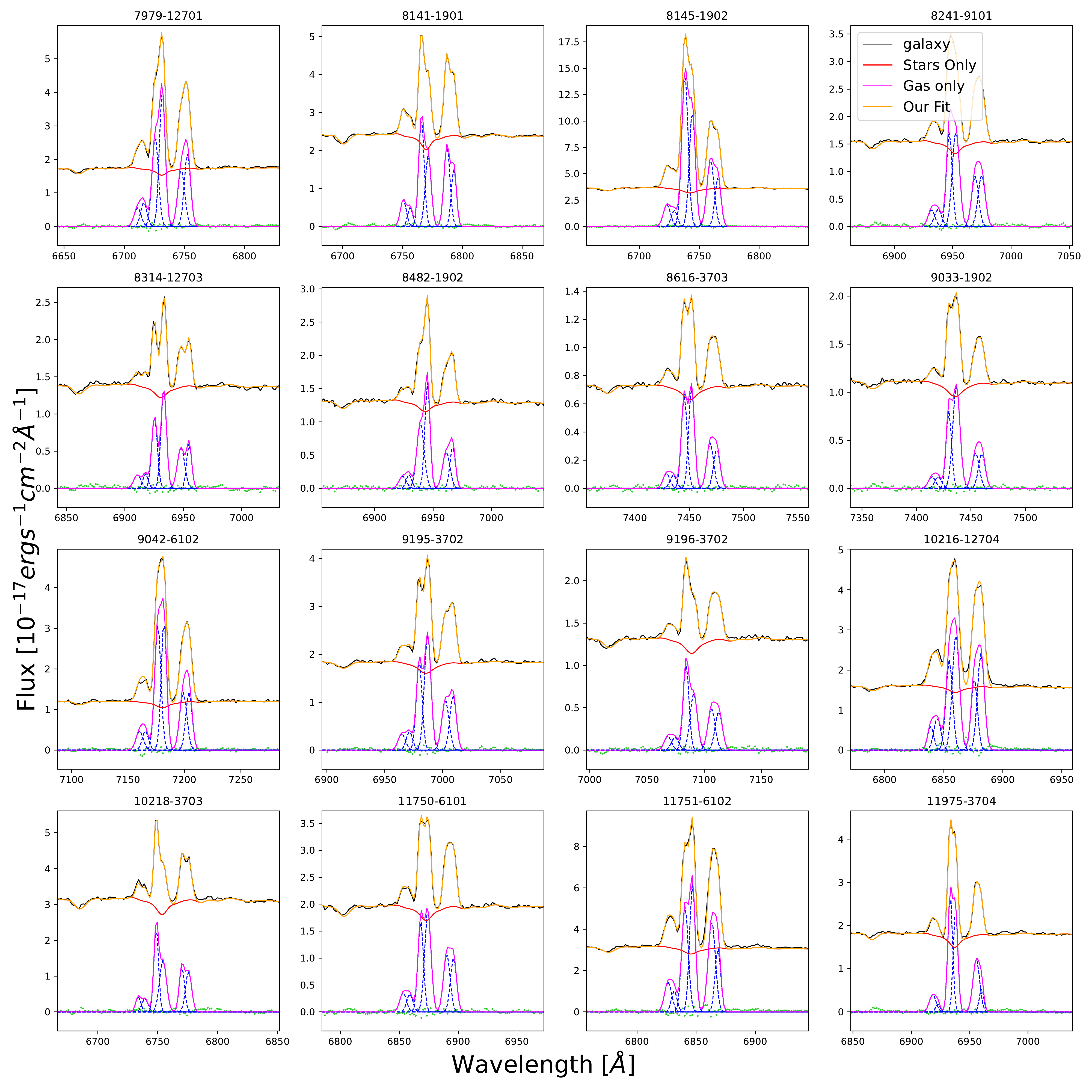}
    \caption{Examples of galaxies with double-peaked narrow lines. Correspondence between spectrum and colour is described in \autoref{fig:egbroadha}. }
    \label{fig:egdualnarrow}
\end{figure*}
\input{Result}
\begin{figure*}
	\includegraphics[width=\textwidth]{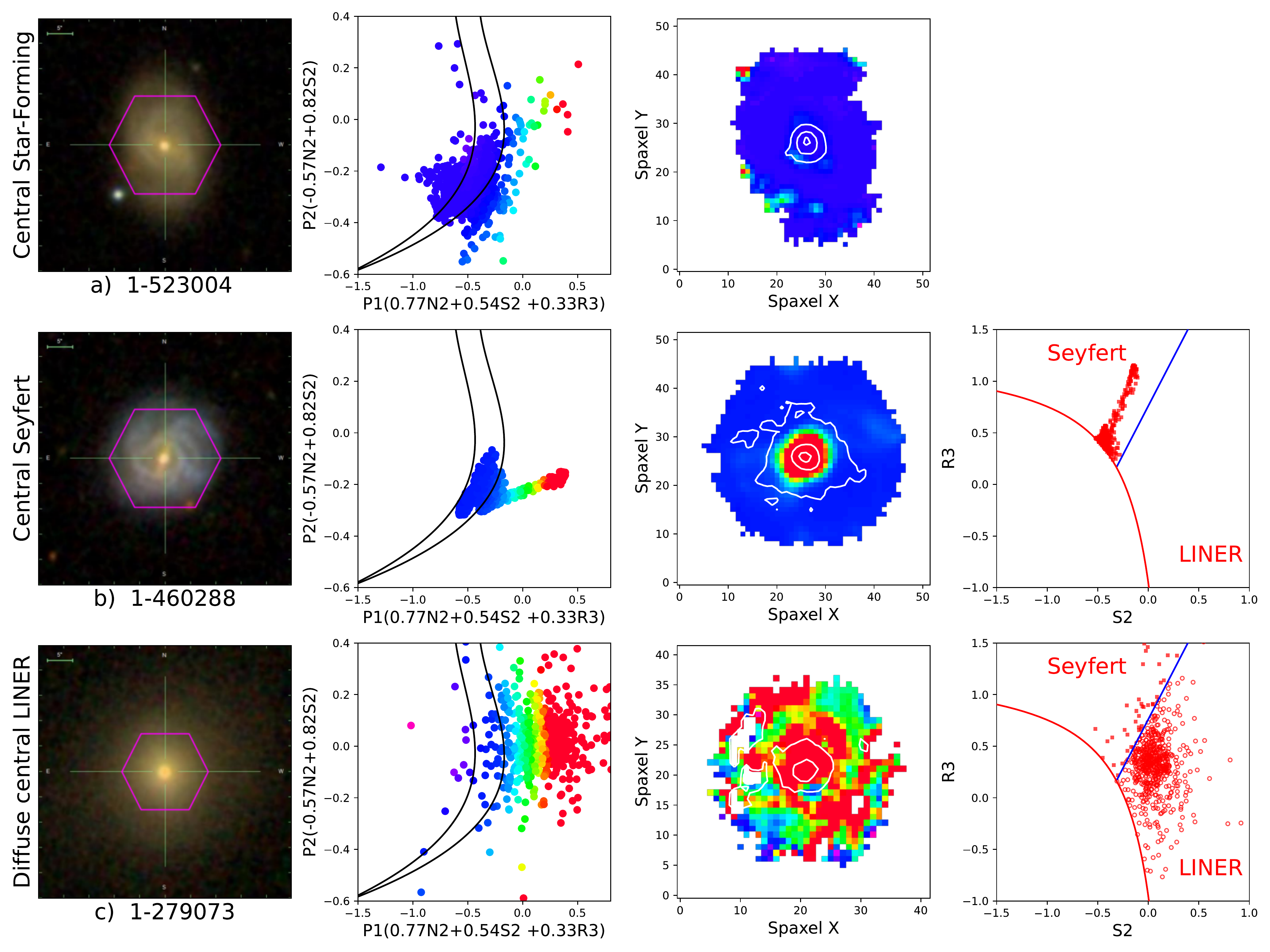}
    \caption{ Examples of the different ionization levels in broad line galaxies. The first column is the galaxy images with MaNGA-ID labeled below. The second column is the classification based on \autoref{BPTclass}. Each dot represents a single spaxel and the colour is based on its distance to $\rm 1\sigma$ line. The two solid black lines are the 1$\sigma$ and 3$\sigma$ limits of the AGN vs star forming separation by \citet[Figure 15]{3DBPT}. The third column is a map of the galaxy colour-coded according to the position of individual regions with the same colour-coding as in the second column. Broad line areas are marked on the map as contours. The outermost contour represents all spaxels with broad $\rm H\alpha$. The innermost contour represents the position of the spaxel with the brightest broad $\rm H\alpha$. The contour in the middle represents the half maximum flux of the broad $\rm H\alpha$. The fourth column separate LINERs (red open circles) from Seyfert AGNs (red filled squares).}
    \label{fig:BPTexample}
\end{figure*}
\section{Results and discussions}
\label{sec:discussion}
\input{discussion}

\section*{Acknowledgements}
We thank the referee for helpful comments that improved the paper. This work (for YF and SM) is partly supported by the National Key Research and Development Program of China (No. 2018YFA0404501), and by the National Nature Science Foundation of China (Grant No. 11821303, 11761131004, and 11761141012) and by the Tsinghua University Initiative Scientific Research Program ID 2019Z07L02017. We also acknowledge the science research grants from the China Manned Space Project with NO. CMS-CSST-2021-A11. 

Funding for the Sloan Digital Sky Survey IV has been provided by the Alfred P. Sloan Foundation, the U.S. Department of Energy Office of Science, and the Participating Institutions. 

SDSS-IV acknowledges support and resources from the Center for High Performance Computing  at the University of Utah. The SDSS website is www.sdss.org.

SDSS-IV is managed by the Astrophysical Research Consortium for the Participating Institutions of the SDSS Collaboration including the Brazilian Participation Group, the Carnegie Institution for Science, Carnegie Mellon University, Center for Astrophysics | Harvard \& Smithsonian, the Chilean Participation Group, the French Participation Group, Instituto de Astrof\'isica de Canarias, The Johns Hopkins University, Kavli Institute for the Physics and Mathematics of the Universe (IPMU) University of Tokyo, the Korean Participation Group, Lawrence Berkeley National Laboratory, Leibniz Institut f\"ur Astrophysik Potsdam (AIP),  Max-Planck-Institut 
f\"ur Astronomie (MPIA Heidelberg), Max-Planck-Institut f\"ur Astrophysik (MPA Garching), Max-Planck-Institut f\"ur Extraterrestrische Physik (MPE), National Astronomical Observatories of China, New Mexico State University, New York University, University of Notre Dame, Observat\'ario Nacional / MCTI, The Ohio State University, Pennsylvania State University, Shanghai Astronomical Observatory, United Kingdom Participation Group, Universidad Nacional Aut\'onoma de M\'exico, University of Arizona, University of Colorado Boulder, University of Oxford, University of Portsmouth, University of Utah, University of Virginia, University of Washington, University of Wisconsin, Vanderbilt University, and Yale University.

\section*{Data Availability}
The data underlying this article are available in the article and in its online supplementary material.


\bibliographystyle{mnras}
\bibliography{example} 




\appendix
\section{The complete catalogue of galaxies with broad line or double narrow line features in MaNGA}
\label{Appendix A}
\onecolumn
\begin{longtable}{llllllll}

  \caption{The complete catalogue of galaxies with broad line features in MaNGA}\\
  
    \hline
    MaNGA-ID & RA & Dec &$\lg L^{\rm broad}_{H\alpha}$ & $\lg\sigma^{\rm broad}_{H\alpha}$ & $\lg\sigma_{\rm e}$ & Extra templates \\
    & (deg) & (deg) & ( $10^{40}$erg s$^{-1}$) & (km s$^{-1}$) & (km s$^{-1}$) &\\
    \hline
    \endfirsthead
    \multicolumn{8}{c}{Table A1 (continued). }\\ 
    \hline
    MaNGA-ID & RA & Dec &$\lg L^{\rm broad}_{H\alpha}$ & $\lg\sigma^{\rm broad}_{H\alpha}$ & $\lg\sigma_{\rm e}$ & Extra templates \\
    & (deg) & (deg) & ( $10^{40}$erg s$^{-1}$) & (km s$^{-1}$) & (km s$^{-1}$) &\\
    \hline
    \endhead
    1-382273 & 118.634242 & 16.809729 & 12.91  & 3.403 & 2.137 & Broad H$\alpha$            \\
    1-547191 & 258.118529 & 35.884086 & 14.71   & 3.251 & 1.954 & Broad Balmer               \\
1-300461 & 137.160233 & 32.592953 & 6.31   & 3.098 & 2.114 & Broad H$\alpha$            \\
1-587259 & 160.228870  & 5.991890   & 12.61   & 2.951 & 1.964 & Broad H$\alpha$,{[}OIII{]} \\
1-150947 & 183.263995 & 51.648597 & 47.03  & 2.958 & 2.332 & Broad H$\alpha$,{[}OIII{]} \\
1-60653  & 150.529327 & 3.057686  & 6.48   & 3.270  & 1.996 & Broad H$\alpha$            \\
1-78719  & 151.006758 & 4.051176  & 4.02   & 2.843 & 1.785 & Broad H$\alpha$            \\
1-3050   & 160.143696 & 0.817429  & 11.29  & 3.364 & 2.137 & Broad H$\alpha$            \\
1-181172 & 140.011912 & 5.735586  & 2.21   & 2.928 & 2.045 & Broad H$\alpha$            \\
1-180998 & 141.915721 & 5.053476  & 38.26  & 2.981 & 2.143 & Broad H$\alpha$,{[}OIII{]} \\
1-244377 & 206.175720  & 55.887093 & 53.05  & 2.925 & 2.358 & Broad H$\alpha$,{[}OIII{]} \\
1-218981 & 127.784888 & 30.895607 & 5.22   & 2.907 & 2.061 & Broad H$\alpha$            \\
1-301330 & 145.829803 & 36.247822 & 2.50    & 2.945 & 2.274 & Broad H$\alpha$,{[}OIII{]} \\
1-301090 & 146.628758 & 34.916838 & 13.15  & 2.841 & 1.892 & Broad H$\alpha$            \\
1-567397 & 145.267159 & 34.732879 & 13.66  & 3.156 & 2.188 & Broad H$\alpha$            \\
1-282637 & 187.964780  & 43.245526 & 2.99   & 2.899 & 2.022 & Broad H$\alpha$            \\
1-195337 & 194.915701 & 53.341186 & 7.01   & 2.953 & 2.312 & Broad H$\alpha$            \\
1-320796 & 214.930149 & 49.236646 & 3.75   & 2.837 & 1.940  & Broad H$\alpha$            \\
1-623700 & 195.169935 & 52.537742 & 10.47  & 2.862 & 1.988  & Broad H$\alpha$,{[}OIII{]} \\
1-135812 & 250.292722 & 38.831981 & 4.49   & 3.023 & 2.001  & Broad H$\alpha$            \\
1-136044 & 252.289916 & 36.223838 & 8.28   & 3.040  & 2.292 & Broad H$\alpha$            \\
1-336095 & 234.082184 & 9.344213  & 1.77   & 2.865 & 1.973 & Broad H$\alpha$,{[}OIII{]} \\
1-294864 & 250.281830  & 22.823568 & 1.57   & 2.913 & 1.820  & Broad H$\alpha$            \\
1-96451  & 311.310818 & -5.484327 & 5.90    & 2.854 & 2.057 & Broad H$\alpha$            \\
1-635845 & 328.577920  & 0.355400    & 8.30    & 2.906 & 1.908 & Broad H$\alpha$,{[}OIII{]} \\
1-53288  & 134.619164 & 0.023460   & 2.64   & 2.908 & 1.914 & Broad H$\alpha$,{[}OIII{]} \\
1-212706 & 325.590823 & -8.364872 & 4.44   & 3.077 & 1.845 & Broad H$\alpha$            \\
1-98401  & 328.108473 & -8.173600   & 13.94  & 2.998 & 2.134 & Broad H$\alpha$,{[}OIII{]} \\
1-113405 & 316.841308 & 11.066421 & 1.71   & 3.247 & 1.903 & Broad H$\alpha$            \\
1-24148  & 258.827410  & 57.658770  & 2.46   & 3.262 & 2.041 & Broad H$\alpha$            \\
1-22948  & 254.542084 & 62.415648 & 16.68  & 2.919 & 2.111 & Broad H$\alpha$,{[}OIII{]} \\
1-109378 & 51.169918  & -0.681473 & 13.74   & 2.850  & 1.914 & Broad H$\alpha$            \\
1-39759  & 12.413884  & 13.685225 & 14.67  & 2.986 & 2.161 & Broad H$\alpha$,{[}OIII{]} \\
1-39615  & 11.876450   & 15.697081 & 3.54   & 3.368 & 2.041 & Broad H$\alpha$            \\
1-548024 & 111.733682 & 41.026691 & 29.12  & 3.410  & 2.212 & Broad H$\alpha$            \\
1-338922 & 114.775747 & 44.402765 & 8.84   & 2.848  & 2.320  & Broad H$\alpha$            \\
1-72322  & 121.014201 & 40.802613 & 29.88  & 3.006 & 2.328 & Broad H$\alpha$,{[}OIII{]} \\
1-121075 & 114.695391 & 29.891283 & 18.77  & 3.064 & 2.188 & Broad H$\alpha$            \\
1-163831 & 118.627843 & 25.815986 & 5.12   & 2.966 & 1.959 & Broad H$\alpha$            \\
1-460288 & 126.059633 & 17.331951 & 6.29   & 2.908 & 1.886 & Broad H$\alpha$,{[}OIII{]} \\
1-460812 & 127.170800   & 17.581400   & 6.68   & 3.303 & 2.220  & Broad H$\alpha$            \\
1-585513 & 138.981378 & 44.332761 & 6.76   & 2.869 & 2.037 & Broad H$\alpha$            \\
1-258373 & 182.286727 & 44.003165 & 5.41   & 2.994 & 1.973 & Broad H$\alpha$,{[}OIII{]} \\
1-558912 & 166.129408 & 42.624554 & 24.44  & 2.874 & 2.375 & Broad H$\alpha$,{[}OIII{]} \\
1-284293 & 197.239319 & 45.905447 & 53.19  & 2.901   & 2.253 & Broad H$\alpha$            \\
1-575726 & 198.958904 & 46.338830  & 6.17   & 2.949 & 2.140  & Broad H$\alpha$            \\
1-261280 & 214.096447 & 38.190986 & 18.18  & 2.898 & 2.097 & Broad H$\alpha$,{[}OIII{]} \\
1-155975 & 142.778168 & 49.079746 & 6.17   & 3.020  & 2.083 & Broad H$\alpha$,{[}OIII{]} \\
1-166919 & 146.709100   & 43.423843 & 9.84   & 3.019 & 2.149 & Broad H$\alpha$            \\
1-211311 & 248.426386 & 39.185120  & 11.82   & 2.932 & 1.934 & Broad H$\alpha$            \\
1-576292 & 225.691343 & 53.512170  & 4.31   & 2.970  & 2.117 & Broad H$\alpha$            \\
1-198182 & 224.749647 & 48.409855 & 5.32   & 2.936 & 2.316 & Broad H$\alpha$            \\
1-569169 & 247.048171 & 39.821898 & 2.29   & 2.934 & 2.025 & Broad H$\alpha$            \\
1-594493 & 247.159333 & 39.551266 & 8.56   & 3.059 & 2.587 & Broad H$\alpha$            \\
1-95585  & 255.029870  & 37.839502 & 7.82   & 3.387 & 2.029 & Broad H$\alpha$            \\
1-546819 & 257.001371 & 36.344421 & 3.48   & 3.082 & 2.336 & Broad H$\alpha$            \\
1-114820 & 330.423500   & 11.856788 & 2.46   & 2.883 & 2.260  & Broad H$\alpha$            \\
1-24142  & 258.845726 & 57.411184 & 3.38   & 3.189 & 2.471 & Broad H$\alpha$            \\
1-604860 & 118.184153 & 45.949276 & 13.21  & 3.477 & 2.026 & Broad H$\alpha$,{[}OIII{]} \\
1-604907 & 119.920672 & 50.839973 & 92.44  & 2.941 & 2.155 & Broad H$\alpha$,{[}OIII{]} \\
1-44303  & 119.182152 & 44.856709 & 4.02   & 3.314 & 1.869 & Broad H$\alpha$            \\
1-44379  & 120.700706 & 45.034554 & 3.24   & 2.894 & 1.996 & Broad H$\alpha$            \\
1-47409  & 130.407776 & 54.918571 & 8.95   & 3.077 & 2.111 & Broad H$\alpha$            \\
1-574519 & 127.178094 & 45.742555 & 5.09   & 3.129 & 2.013 & Broad H$\alpha$            \\
1-163966 & 120.087418 & 26.613527 & 7.84   & 3.339 & 2.124 & Broad H$\alpha$            \\
1-279073 & 170.588145 & 46.430504 & 11.47   & 2.921 & 2.346 & Broad H$\alpha$            \\
1-173958 & 167.306020  & 49.519432 & 4.33   & 2.855 & 1.908 & Broad H$\alpha$            \\
1-458092 & 203.190094 & 26.580376 & 24.05   & 2.841 & 1.716 & Broad H$\alpha$,{[}OIII{]} \\
1-590142 & 186.908558 & 40.160489 & 47.12   & 2.859 & 2.079 & Broad H$\alpha$            \\
1-173641 & 163.663838 & 47.862282 & 57.45   & 3.218 & 1.949 & Broad H$\alpha$            \\
1-314409 & 223.867459 & 32.840028 & 9.18   & 2.902 & 2.061 & Broad H$\alpha$,{[}OIII{]} \\
1-268479 & 240.475078 & 31.892062 & 54.14  & 3.150  & 2.241 & Broad H$\alpha$            \\
1-210186 & 241.150985 & 43.879788 & 14.76   & 3.062 & 1.869 & Broad H$\alpha$            \\
1-269632 & 247.560973 & 26.206474 & 17.02  & 2.914 & 2.114 & Broad H$\alpha$,{[}OIII{]} \\
1-296733 & 241.717361 & 27.927542 & 8.18   & 3.118 & 2.053 & Broad H$\alpha$,{[}OIII{]} \\
1-265338 & 239.752752 & 27.985371 & 22.93   & 2.999 & 2.057 & Broad H$\alpha$,{[}OIII{]} \\
1-265404 & 239.710279 & 27.390083 & 35.56  & 2.898 & 2.305 & Broad H$\alpha$            \\
1-71872  & 119.617128 & 37.786625 & 33.09   & 3.053 & 2.420  & Broad H$\alpha$            \\
1-71987  & 119.486337 & 39.993365 & 26.30    & 2.894 & 2.072 & Broad H$\alpha$            \\
1-122304 & 121.920806 & 39.004239 & 37.51   & 2.901 & 2.017 & Broad H$\alpha$,{[}OIII{]} \\
1-96589  & 311.938568 & -5.421022 & 13.33   & 3.290  & 2.045 & Broad H$\alpha$            \\
1-37385  & 46.294197  & -1.075470  & 10.65  & 3.401   & 2.134 & Broad H$\alpha$            \\
1-37336  & 46.717069  & -0.896545 & 28.11   & 2.998 & 2.061 & Broad H$\alpha$            \\
1-37633  & 47.142989  & 0.550937  & 42.03   & 3.233 & 2.185 & Broad H$\alpha$            \\
1-603039 & 29.052229  & 14.906639 & 13.25   & 2.904 & 1.929 & Broad H$\alpha$,{[}OIII{]} \\
1-24660  & 262.399283 & 54.494424 & 9.98     & 3.315 & 2.083 & Broad H$\alpha$            \\
1-574506 & 123.330544 & 46.147157 & 8.87   & 3.238 & 2.072 & Broad H$\alpha$            \\
1-413061 & 130.187087 & 22.294456 & 17.84  & 3.477 & 2.121 & Broad H$\alpha$,{[}OIII{]} \\
1-385099 & 129.545574 & 24.895295 & 11.60   & 3.054 & 2.201 & Broad H$\alpha$,{[}OIII{]} \\
1-61630  & 184.029549 & 50.825061 & 57.94 & 3.047 & -- & Broad H$\alpha$              \\
1-113712 & 319.193099 & 11.043741 & 111.18 & 3.217 & 2.493 & Broad Balmer               \\
1-179024 & 312.923122 & 0.859873  & 135.12 & 3.396 & 2.199 & Broad Balmer               \\
1-180204 & 322.213311 & -1.070118 & 33.39   & 3.033 & 1.813 & Broad Balmer               \\
1-596598 & 331.122900   & 12.442626 & 9.39   & 2.989 & 1.987 & Broad Balmer               \\
1-115875 & 338.410077 & 13.212127 & 154.21 & 3.327 & 2.373 & Broad Balmer               \\
1-39766  & 11.830822  & 14.703485 & 15.68  & 3.242 & 2.041 & Broad Balmer, {[}OIII{]}   \\
1-52660  & 61.453262  & -6.323837 & 9.62   & 3.431 & 2.037 & Broad Balmer               \\
1-495383 & 179.294527 & 22.296169 & 23.40   & 3.156 & 2.146 & Broad Balmer               \\
1-235576 & 215.017907 & 47.121330  & 32.41  & 3.084 & 2.093 & Broad Balmer               \\
1-620993 & 189.213253 & 45.651170  & 5.55   & 3.040  & 1.813 & Broad Balmer, {[}OIII{]}   \\
1-418023 & 205.753337 & 36.165656 & 12.88   & 2.985 & 1.806 & Broad Balmer               \\
1-256832 & 169.513447 & 45.113029 & 74.07  & 3.496 & 2.474 & Broad Balmer               \\
1-134271 & 240.214209 & 46.481412 & 3.91   & 3.157 & 1.875 & Broad Balmer               \\
1-210017 & 241.271447 & 45.442992 & 24.95  & 3.329 & 2.076 & Broad Balmer               \\
1-90231  & 234.541844 & 57.603653 & 17.49  & 3.095 & 1.934 & Broad Balmer               \\
1-576315 & 241.325410  & 52.120151 & 5.21   & 3.157 & 2.045 & Broad Balmer               \\
1-631871 & 226.937461 & 51.452832 & 37.97  & 3.131 & 2.170  & Broad Balmer               \\
1-550901 & 321.007912 & -0.366324 & 21.21  & 3.073 & 1.991 & Broad Balmer               \\
1-635646 & 323.115930  & 10.138620  & 602.46 & 3.197 & -- & Broad Balmer               \\
1-378688 & 117.966207 & 49.814311 & 14.12  & 3.099 & 1.959 & Broad Balmer               \\
1-71974  & 118.855394 & 39.186094 & 13.58  & 2.868 & 1.987 & Broad Balmer               \\
1-423024 & 206.007950  & 25.941199 & 14.20   & 2.875 & 2.021 & Broad Balmer               \\
1-149561 & 171.657262 & 51.573041 & 1.67   & 2.940  & 1.806 & Broad Balmer               \\
1-614567 & 171.400654 & 54.382574 & 22.87  & 3.106 & 2.217 & Broad Balmer               \\
1-295542 & 246.255977 & 24.263156 & 18.57  & 3.084 & 2.053 & Broad Balmer               \\
1-633584 & 241.231428 & 28.165759 & 64.84  & 3.233 & 2.182 & Broad Balmer               \\
1-547548 & 260.666382 & 30.881270  & 83.26  & 3.545 & 2.233 & Broad Balmer               \\
1-37863  & 46.664913  & 0.061977  & 174.90  & 3.047 & 2.384 & Broad Balmer               \\
1-574504 & 123.820326 & 46.075253 & 28.03  & 2.841 & 2.149 & Broad Balmer               \\
1-385623 & 131.725411 & 25.370089 & 46.02  & 3.359 & 2.219 & Broad Balmer               \\
1-42214  & 33.239981  & 14.102849 & 27.23  & 3.110  & 2.196 & Broad Balmer               \\
1-50537  & 42.984877  & -8.377050  & 24.99   & 3.010  & 1.903 & Broad Balmer               \\
1-352023 & 124.327367 & 52.029915 & 15.86   & 2.892 & 1.813 & Broad Balmer               \\
1-585230 & 129.747019 & 26.136991 & 18.98  & 2.968 & 2.149 & Broad Balmer               \\
1-607451 & 152.680678 & 6.200391  & 100.59 & 3.076 & 2.165 & Broad Balmer               \\
1-560065 & 141.347425 & 4.509069  & 16.28  & 3.022 & 2.233 & Broad Balmer               \\
1-197677 & 218.718594 & 48.661891 & 28.03  & 3.249 & 2.146 & Broad Balmer               \\
1-175853 & 184.029549 & 50.825062 & 64.71  & 3.283 & 2.139 & Broad Balmer               \\
1-200510 & 243.256777 & 37.287475 & 105.19 & 3.286 & 2.152 & Broad Balmer               \\
1-28725  & 347.334456 & 0.756509  & 14.99  & 3.339 & 1.978 & Broad Balmer               \\
1-597772 & 348.418745 & 14.020977 & 29.33  & 3.064 & 1.991 & Broad Balmer               \\
1-53412  & 134.649946 & 1.530434  & 39.54  & 3.093 & 2.041 & Broad Balmer               \\
1-66643  & 198.274200   & 1.465524  & 28.6   & 2.969 & 1.919 & Broad Balmer               \\
1-568584 & 205.282731 & 23.282055 & 10.25  & 2.923 & 2.033 & Broad Balmer, {[}OIII{]}   \\
1-90242  & 233.968343 & 57.902637 & 138.02 & 3.258 & -- & Broad Balmer               \\
\hline
\label{tab:catalog}\\
\end{longtable}
\begin{tablenotes}
  \item Note: Column (1) is the MaNGA identification ID. Column (2-3) is the position of the galaxy shown in celestial coordinates. We use the positions to acquire optical and radio images. Column (4) is the apparent luminosity ratio measured through the double $\rm H\alpha$ components. Column (5) is the velocity difference between the two narrow components. Column (6) is the stellar velocity dispersion within 1Re corrected by masking the broad line region in \S 3.2. Column (7) shows the extra templates added when fitting each galaxy. The label 'Broad H$\alpha$' means in the broad Balmer set of templates the broad $\rm H\delta$, $\rm H\gamma$, and $\rm H\beta$ components are all zero or below the noise level. Only broad $\rm H\alpha$ is observable. 
  \end{tablenotes}

\begin{longtable}{llllllll}
  \caption{The complete catalogue of galaxies with double narrow line features in MaNGA}\\
    \hline
    MaNGA-ID & Ra & Dec &Luminosity Ratio&  $\lg \Delta V$ (km/s) & $\lg \sigma_{e}$ (km/s) & Extra templates \\
    \hline
    \endfirsthead
    \multicolumn{8}{c}{Table A2 (continued). }\\ 
    \hline
    MaNGA-ID & Ra & Dec &Luminosity Ratio&  $\lg \Delta V$ (km/s) & $\lg \sigma_{e}$ (km/s) & Extra templates \\
    \hline
    \endhead
1-114955 & 332.602090  & 11.713077 & 1.91   & 2.444 & 2.433 & Dual narrow                \\
1-596678 & 332.892838 & 11.795929 & 0.53   & 2.358 & 2.130  & Dual narrow                \\
1-115365 & 333.483204 & 13.755396 & 1.10    & 2.391 & 1.982 & Dual narrow                \\
1-42007  & 33.626933  & 13.257206 & 1.61   & 2.367 & 2.057 & Dual narrow                \\
1-41752  & 31.953063  & 13.609595 & 3.01   & 2.330  & 2.233 & Dual narrow                \\
1-574402 & 115.368720  & 44.408794 & 0.62   & 2.408 & 2.130  & Dual narrow                \\
1-339094 & 117.472421 & 45.248483 & 0.64   & 2.408 & 2.025 & Dual narrow                \\
1-556501 & 117.045741 & 28.230275 & 0.71   & 2.439 & 2.170  & Dual narrow                \\
1-460840 & 126.898019 & 17.918509 & 1.25   & 2.391 & 2.124 & Dual narrow                \\
1-278057 & 166.767439 & 45.822139 & 0.48   & 2.371 & 2.127 & Dual narrow                \\
1-258380 & 181.545971 & 45.149206 & 0.66   & 2.360  & 2.182 & Dual narrow                \\
1-248638 & 243.212322 & 40.318839 & 1.39   & 2.524 & 1.954 & Dual narrow                \\
1-251269 & 209.163574 & 43.585641 & 1.30    & 2.344 & 2.114 & Dual narrow                \\
1-92547  & 243.102565 & 48.529886 & 1.11   & 2.405 & 2.158 & Dual narrow                \\
1-542392 & 244.216717 & 50.282196 & 0.89   & 2.378 & 2.017 & Dual narrow                \\
1-593159 & 217.629971 & 52.707159 & 2.39   & 2.465 & 2.107 & Dual narrow                \\
1-547744 & 155.251009 & 36.151845 & 0.64   & 2.386 & 2.083 & Dual narrow                \\
1-631278 & 221.437984 & 51.580820  & 0.41   & 2.360  & 2.196 & Dual narrow                \\
1-180308 & 322.510226 & 0.464199  & 1.01   & 2.431 & 2.190  & Dual narrow                \\
1-190574 & 311.780971 & 0.300461  & 0.96   & 2.312 & 2.149 & Dual narrow                \\
1-626502 & 203.057063 & 26.949981 & 0.88   & 2.362 & 1.968 & Dual narrow                \\
1-149536 & 172.104500   & 51.028353 & 0.80    & 2.250  & 2.049 & Dual narrow                \\
1-321823 & 223.218540  & 45.234801 & 1.87   & 2.439 & 2.267 & Dual narrow                \\
1-375053 & 233.840978 & 27.124640  & 1.06   & 2.356 & 2.143 & Dual narrow                \\
1-316872 & 235.152679 & 28.512436 & 1.21   & 2.314 & 2.029 & Dual narrow                \\
1-376413 & 241.913639 & 23.416866 & 1.11   & 2.389 & 2.093 & Dual narrow                \\
1-179060 & 313.323377 & 0.120355  & 1.47   & 2.393 & 2.176 & Dual narrow                \\
1-633683 & 241.866219 & 27.569766 & 1.95   & 2.480  & 2.107 & Dual narrow                \\
1-42250  & 27.842784  & 13.060335 & 1.56   & 2.490  & 2.241 & Dual narrow                \\
1-24641  & 261.927878 & 54.052347 & 0.86   & 2.446 & 2.220  & Dual narrow                \\
1-384945 & 129.147411 & 23.119557 & 0.69   & 2.367 & 2.185 & Dual narrow                \\
1-556749 & 120.016894 & 23.437849 & 1.06   & 2.572 & 2.167 & Dual narrow                \\
1-604912 & 120.082382 & 26.701442 & 1.01   & 2.423 & 2.223 & Dual narrow                \\
1-556492 & 121.701226 & 28.421494 & 1.12   & 2.471 & 2.223 & Dual narrow                \\
1-382452 & 118.162300   & 18.321606 & 1.54   & 2.389 & 2.241 & Dual narrow                \\
1-383171 & 119.878652 & 17.491422 & 0.95   & 2.396 & 2.093 & Dual narrow                \\
1-298623 & 123.098008 & 24.115978 & 0.03   & 2.619 & 2.050   & Dual narrow                \\
1-300318 & 135.250303 & 31.090696 & 1.03   & 2.330  & 2.072 & Dual narrow                \\
1-214286 & 132.177854 & 3.806221  & 1.09   & 2.591 & 2.384 & Dual narrow                \\
1-605611 & 136.015437 & 3.584720   & 0.50    & 2.340  & 2.041 & Dual narrow                \\
1-319646 & 203.674288 & 50.464168 & 0.34   & 2.350  & 2.072 & Dual narrow                \\
1-319387 & 200.971920  & 51.435235 & 0.46   & 2.292 & 2.013 & Dual narrow                \\
1-397168 & 188.985416 & 39.836432 & 1.18   & 2.417 & 2.158 & Dual narrow                \\
1-301470 & 146.633472 & 35.330342 & 0.73   & 2.405 & 2.185 & Dual narrow                \\
1-319387 & 200.971920  & 51.435235 & 0.76   & 2.297 & 2.017 & Dual narrow                \\
1-270445 & 251.890170  & 25.118840  & 0.80    & 2.297 & 2.111 & Dual narrow                \\
1-246331 & 221.786519 & 50.195707 & 0.93   & 2.270  & 1.978 & Dual narrow               \\
1-77704 & 232.406655 & 37.786321 & 1.14 & 2.331 & 2.176 & Dual narrow               \\
1-22948 & 254.542084 & 62.415648 & 0.63 & 2.293 & 2.195 & Dual narrow               \\

\hline

  \label{tab:catalog2}\\
\end{longtable}
\begin{tablenotes}
  \item Note: Column (1) is the MaNGA identification ID. Column (2-3) is the position of the galaxy shown in celestial coordinates. We use the positions to acquire optical and radio images. Column (4) is the apparent luminosity ratio measured through the double $\rm H\alpha$ components. Column (5) is the velocity difference between the two narrow components. Column (6) is the stellar velocity dispersion within 1Re. Column (7) shows the extra templates added when fitting each galaxy. The label \texttt{`Dual narrow'} means adding an extra set of narrow lines with kinematic constrains from \autoref{tab:emiconstrain3}. 
\end{tablenotes}

\bsp	
\label{lastpage}

\end{document}

%% file: introduction.tex
The emission-line properties and kinematics of gas components in galaxies is a valuable probe to understand the dynamic properties and the formation and evolution of galaxies. Certain dynamic processes such as the biconical outflow of gas or feedback processes of active galactic nuclei (AGNs) exhibit characteristic spectral features across the optical spectrum. \HL{A large number of biconical outflows of gas are found driven by AGNs \citep{dnarrow1,dnarrow2,dnarrow3,dnarrow4,Lopez-Coba_2020} or star formation \citep{dnarrowSF1,dnarrowSF2,Lopez-Coba_2019,Lopez-Coba_2020} and are correlated with the star-formation rate (SFR)  density \citep{Lopez-Coba_2019}. } They are usually characterized by its double-peaked narrow emission lines in spectrum while type-1 AGNs show broad and narrow emission lines in their central nuclear region \citep{clftypeIAGN1,clftypeIAGN2}. 

Much efforts have been made in this field to acquire catalogs of galaxies with multi-Gaussian emission line features. For example, \cite{2015_Oh} searched for the presence of broad Balmer emission lines in SDSS DR7 galaxy database. The Swift BAT AGN
Spectroscopic Survey (BASS) \citep{Swift-BAT} presents an AGN catalog in DR1 \citep{Baumgartner_2013_BASS_DR1,Koss_2017_BASS_DR1} and DR2 \citep{BASS_AGN_DR2}, using broad Balmer lines fits to classify type-1 AGNs. \HL{\cite{Lacerda_2020} present an optically-selected AGN catalog of 34 AGNs within a sample of 867 galaxies extracted from the extended Calar-Alto Legacy Integral Field spectroscopy Area (eCALIFA) survey \citep{2016c,Galbany_2018}.} \cite{dnarrow1,dnarrow3} used catalogs of double-peaked narrow lines to search for evidence of dual AGNs and feedback from biconical AGN outflows. More recently \cite{dnarrowSF2} used both features to study biconical outflows dirven by star-formation process. 

The Sloan Digital Sky Survey-IV (SDSS-IV) Mapping Nearby Galaxies at Apache Point Observatory \citep[MaNGA,][]{overviemanga} integral field unit (IFU) survey provides spatially resolved spectra for $\sim$10000 nearby galaxies. Gas kinematics and emission-line properties in every spaxel of the galaxies are produced by the Data Analysis Pipeline (DAP) \citep{DAP,DAPemissionline}. DAP derives the kinematic information from the IFU spectra of galaxies by fitting emission and absorption lines using the {\sc ppxf} software \citep{ppxf2004,pPXF,Cappellari2022} with the combination of the MILES stellar library \citep{MILESlibrary,MILESlibrary2}. The current DAP assumes a single Gaussian component per emission line \citep{DAPemissionline}, which represents a notable limitation. It is not currently possible to recover either the broad and narrow Balmer line features of type-1 AGN or the double-peaked narrow emission lines of gas outflows since these cannot be well approximated by single Gaussian emission lines. Previously, \cite{Sanchez_2018,Sanchez_2022} presented a catalog of type-1 and type-2 AGNs based on emission line ratios and $\rm H\alpha$ equivalent width in MaNGA galaxies. However, their selection of type-1 AGNs is limited to multi-Gaussian fitting only within the wavelength range covered by $\rm H\alpha$ and the [NII] doublet. \cite{cor_2022} analysed the multiwavelength properties of type-1 AGN in DR15 sample. Building on these previous works, we further provide accurate measurements of broad line properties and a more complete catalog of broad and double-peaked emission line galaxies in MANGA for further multi-wavelength analysis. 

A precise determination of the properties of multi-Gaussian component is critical to understand the peculiar dynamic processes behind it. The velocity dispersion of a broad $\rm H\alpha$ line is important for studying galaxy evolution, black hole activity and feedback processes of type-1 AGNs \citep{M-broad,feedback2007_BPT5,broad2011,broad2018,broad2019,feedback2018_BPT6,AGNkenimatics}. The velocities and velocity difference of double-peaked narrow emission lines play an important role in kinematic classifications of biconical AGN outflows and AGN feedback processes \citep{dnarrow1,dnarrow2,dnarrow3}. Here we report an effort to determine the multi-Gaussian emission line parameters for MaNGA spectra in which a single Gaussian approximation fails. 

This paper is organized as follows. In \autoref{sec:sample}, we describe the selection criteria whereby we find our sample of galaxies with possible multi-Gaussian features in MaNGA dataset. In \autoref{sec:methods}, we demonstrate how we use multi-Gaussian templates based on {\sc ppxf} to fit these galaxies and present the catalog of galaxies with broad $\rm H\alpha$ or dual narrow emission line features. Finally, we discuss our findings in \autoref{sec:discussion}.

%% file: Sample.tex
This section shows how we use the DAP residual to select the spectra where DAP fails to fit precisely with a single Gaussian.

\begin{figure*}
	\includegraphics[width=\textwidth]{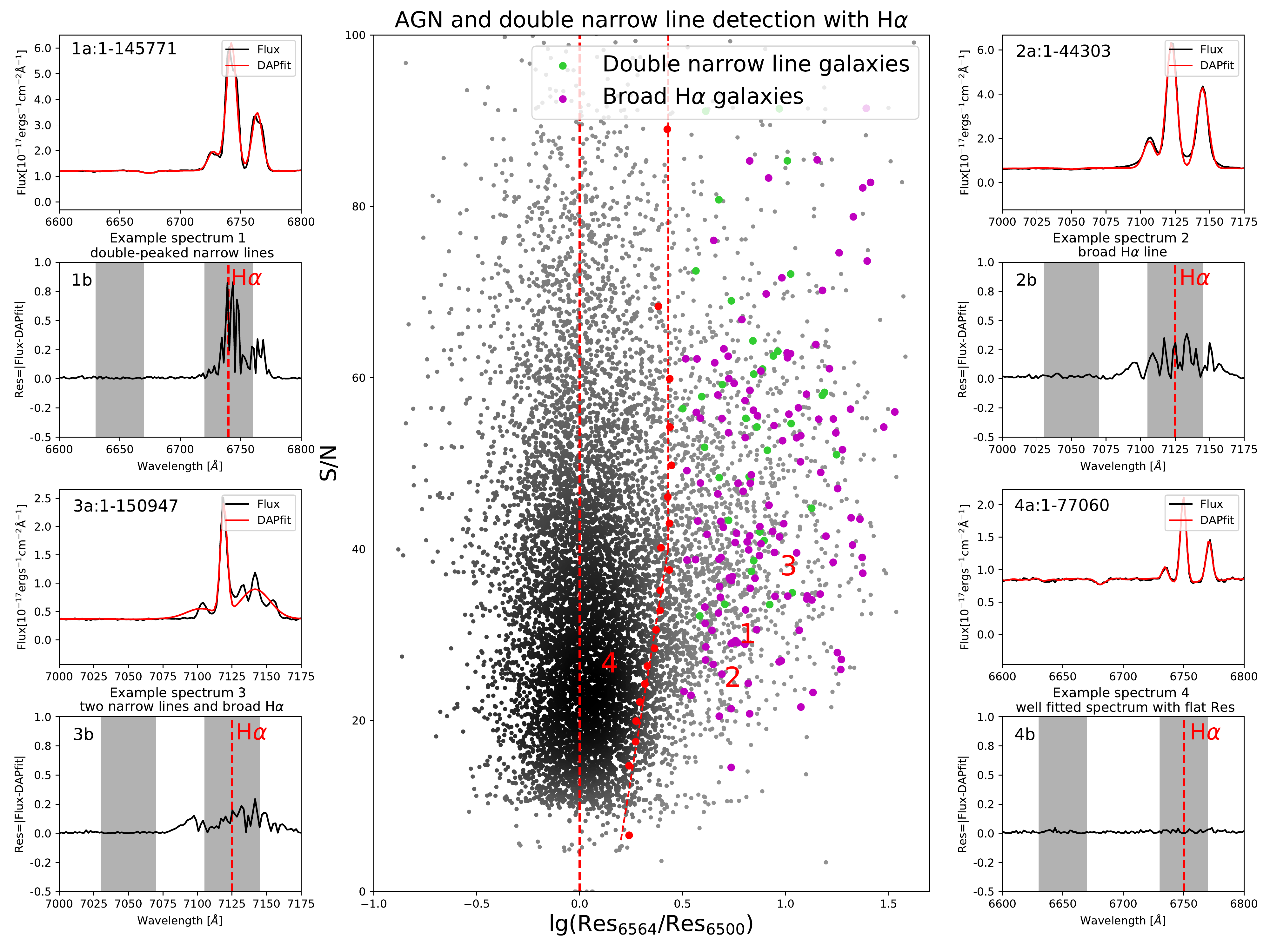}
    \caption{Normalized residual distribution vs. mean g-band signal-to-noise-ratio. The red dots represent 1$\sigma$ cuts of normalized residual in 20 S/N bins, with the red dashed line indicating the trend of the division. The coloured dots shows how the broad $\rm H\alpha$ and double-peaked narrow line galaxies are distributed according to our results in \autoref{sec:fitresult}. The eight adjacent panels are from four different galaxies and demonstrate three dominant multi-Gaussian features: 1) double-peaked narrow $\rm H\alpha$ line, 2) broad $\rm H\alpha$ line, 3) double-peaked narrow $\rm H\alpha$ line and broad line, and 4) a well-fitted spectrum. The flux (a) and residuals (b) are shown two in a group. The shaded part of spectra indicates the area affected by the $\rm H\alpha$ line and area used to calculate $\rm Res_{6500}$. }
    \label{fig:Hares}
\end{figure*}
\begin{figure*}
	\includegraphics[width=\textwidth]{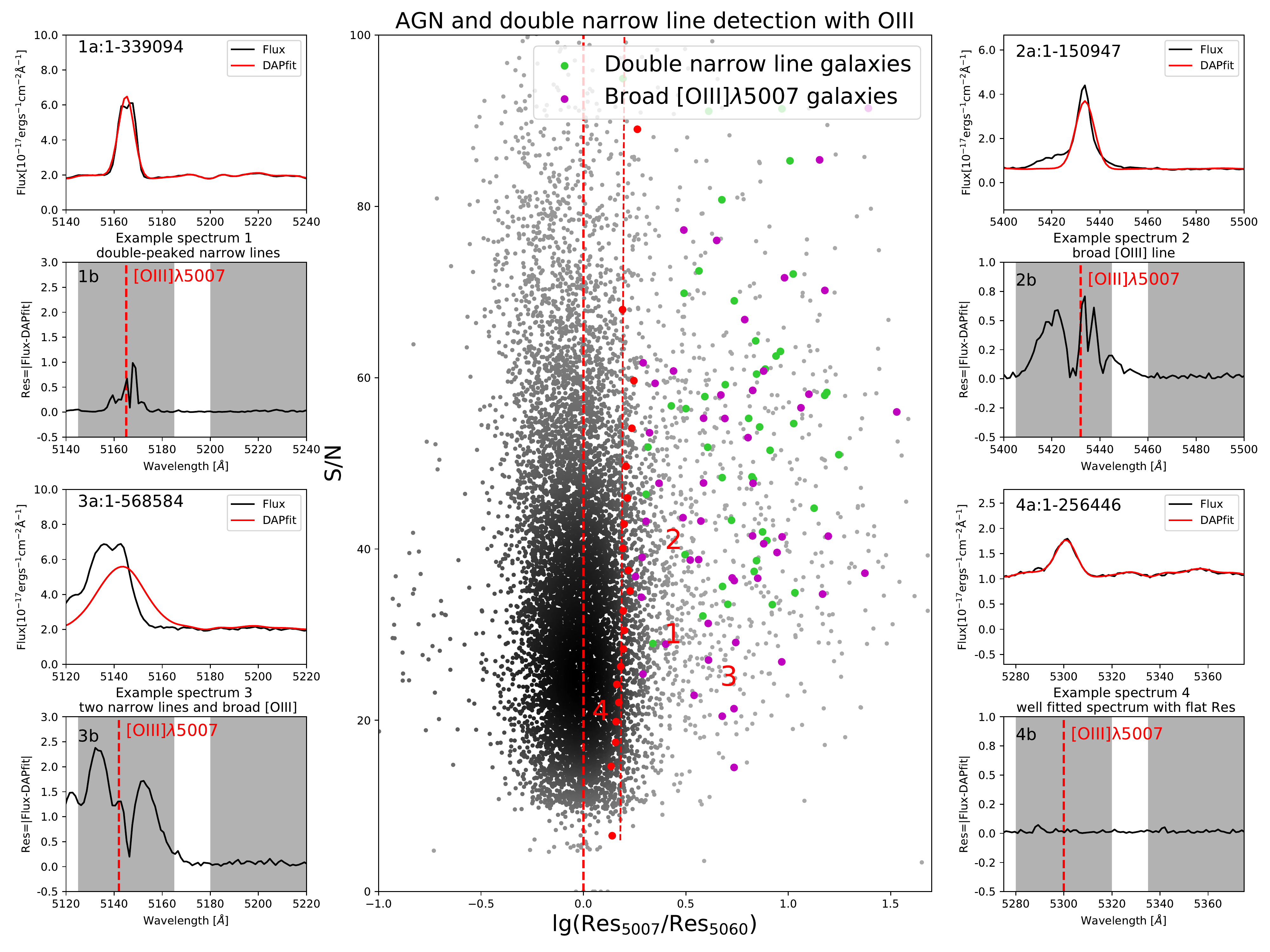}
    \caption{Normalized residual distribution vs. mean g-band signal-to-noise-ratio. The red dots represent 1$\sigma$ cuts of normalized residual in 20 ($S/N$) bins, with the red dashed line indicating the trend of the division. The coloured dots shows how the broad [OIII]$\rm \lambda$5007 and double-peaked narrow line galaxies are distributed according to our results in \autoref{sec:fitresult}. It should be noted that the double-peaked narrow line galaxies are the same galaxies shown in  \autoref{fig:Hares} because double-peaked narrow line feature appears in every emission line. The eight adjacent panels are from four different galaxies and demonstrate three dominant multi-Gaussian features: 1) double-peaked narrow [OIII]$\rm \lambda$5007 line, 2) broad [OIII]$\rm \lambda$5007 line, 3) double-peaked narrow [OIII]$\rm \lambda$5007 line and broad line, and 4) a well-fitted spectrum. The flux (a) and residuals (b) are shown two in a group. The shaded part of spectra indicates the area affected by the [OIII]$\rm \lambda$5007 line and area used to calculate $\rm Res_{5060}$. }
    \label{fig:OIIIres}
\end{figure*}

\subsection{The MaNGA Survey}
The MaNGA survey \citep{overviemanga} is an integral field unit (IFU) survey providing a three-dimensional view of galaxies. MaNGA data includes a  sample of 10010 high-quality unique galaxy observations with spatially resolved spectra within the redshift range of 0.01 < z < 0.15 \citep{wake2017}. The spaxel size of MaNGA is 0.5" and the average g-band Point Spread Function (PSF) FWHM (full width at half maximum) throughout the survey is about 2.54" \citep{DRP}.

The spectra provided by MaNGA cover a wavelength range of 3600 to 10300 \AA \, at a spectral resolution of $\rm \sigma$ = 72 km/s \citep{DRP}. Raw spectra are processed by the Data Reduction Pipeline (DRP) and products such as stellar continuum and emission line properties are produced by the Data Analysis Pipline (DAP) \citep{DRP,DAP,DAPemissionline}. 
\subsection{Selection Based on Normalized Residual}
In total, we have 10010 high-quality unique galaxy observations from DR17, the final release of MaNGA, with MaNGA DAP outputs. An empirical way of assessing line detection is using the amplitude over noise ratio ($(A/N)_{\rm line}>3$) acquired from the DAP emission line modeling procedure \citep{DAPemissionline}. It is safe to remove 293 galaxies with a maximum $(A/N)_{\rm line}<3$ because there are no emission line detection in these galaxies. That leaves only 9717 galaxies to measure the fit quality of the emission line model. 

A DAP residual is introduced to represent the fit quality of the spectrum. The DAP residual is defined as the absolute value of the difference between the DAP model and the actual flux data:
\begin{equation}
  \rm Residual=|DAP Model - Flux|.
  \label{eq:Residual}
\end{equation}

The residual of a given wavelength is calculated from the median residual within a range of $\pm$20\AA \, around the wavelength. The emission line residual is normalized by comparing it with the residual at an adjacent wavelength in the same spectrum without emission lines. We use the normalized residual of two typical emission lines, $\rm H\alpha$ and [OIII]$\rm \lambda$5007, to represent the fit quality of emission lines. $\rm H\alpha$ is the strongest emission line in the Balmer series and the broad $\rm H\alpha$ line is the most prominent feature of type-1 AGNs while [OIII] $\rm \lambda$5007 is usually a relatively strong forbidden line in 4000\AA $\sim$ 8000\AA \, with no contamination from other emission lines close to it. The normalized residual of $\rm H{\alpha}$ (6564\AA) is calculated through \autoref{eq:NRHa} and the normalized residual of [OIII]$\rm \lambda$5007 is calculated through \autoref{eq:NROIII}.

\begin{equation}
  \frac{\rm Res_{6564}}{\rm Res_{6500}}  =\frac{{\rm Median\, Residuals\, within}\, (1+z)\times [6544,6584] \mbox{\AA}  }{{\rm Median\, Residuals\, within}\, (1+z)\times [6480,6520] \mbox{\AA}}.
  \label{eq:NRHa} 
\end{equation}

\begin{equation}
  \frac{\rm Res_{5007}}{\rm Res_{5060}}
 =\frac{{\rm Median\, Residuals\, within}\,  (1+z)\times [4987,5027] \mbox{\AA}  }{{\rm Median\, Residuals\, within}\,  (1+z)\times [5040,5080] \mbox{\AA}}.
  \label{eq:NROIII} 
\end{equation}

Ideally, we will expect the normalized residual to be roughly 1, or slightly larger due to the increased photon noise due to the emission, which means the spectra of the emission lines are as well-fitted as anywhere else in a spectrum. The galaxies are divided uniformly into 20 signal-to-noise ratio ($S/N$) bins according to their mean g-band weighted signal-to-noise ratio. The middle panel in  \autoref{fig:Hares} shows the distribution of normalized residual in galaxies with $(A/N)_{\rm H\alpha} > 3$. The red dots represent the position of 1$\rm \sigma$ cut in each ($S/N$) bin. A total of 17\% of galaxies have a residual ratio above this cut. The ordinate of each point represents the median ($S/N$) in its corresponding bin. The dashed line only represents the division trend. The figure shows that the 1$\sigma$ residual cuts increase with the increase of ($S/N$) for $S/N<40$. That is because brighter galaxies with stronger $\rm H\alpha$ emission have larger ($S/N$). These galaxies have larger residuals within certain ($S/N$) range. The outliers to the right of the red dots are candidates for searching broad $\rm H\alpha$ and double-peaked narrow line features. \par

\autoref{fig:OIIIres} shows the distribution of normalized residual vs. ($S/N$) for [OIII] $\rm \lambda$5007 in galaxies with $(A/N)_{\rm H\alpha} > 3$. The structure and labels of image is consistent with  \autoref{fig:Hares}. It should be noted that the number of outliers in the [OIII] $\rm \lambda$5007 region is much smaller than that of the former $\rm H\alpha$ outliers. That is because only a fraction of type-1 AGNs shows broad [OIII] feature. 

We select the galaxies to the right of the 1$\rm \sigma$ cuts in  \autoref{fig:Hares} and \autoref{fig:OIIIres}. There are 1652 different galaxies. The flowchart in \autoref{fig:flow} shows the number of objects that survived each selection. 

\begin{figure}
\begin{tikzpicture}[node distance=1.2cm]
    \node (start) [rectangle, draw, text width=5cm, align=center] {High-quality unique galaxy observations in MaNGA};
    \node (input) [rectangle, draw, below of=start, text width=5cm, align=center] {$\rm H\alpha$ or $\rm [OIII]\lambda 5007$ emission line with central $A/N>3$};
    \node (output) [rectangle, draw, below of=input, text width=5cm, align=center] {1$\sigma$ outliers from Res vs. ($S/N$) distribution};
    \node (stop) [rectangle, draw, below of=output, text width=5cm, align=center] {139 broad $\rm H\alpha$ emissions including 38 broad OIII doublets, 49 double narrow line};
    
    \draw [->] (start) -- (input);
    \draw [->] (input) -- (output);
    \draw [->] (output) -- (stop);
    
    \node [left of=start, xshift=-2cm, yshift=0cm] {N=10010};
    \node [left of=input, xshift=-2cm, yshift=0cm] {N=9717};
    \node [left of=output, xshift=-2cm, yshift=0cm] {N=1652};
    \node [left of=stop, xshift=-2cm, yshift=0cm] {N=188};

\end{tikzpicture}
\caption{Number of objects survived after each selection step.}
\label{fig:flow}
\end{figure}

%% file: Result.tex
In this section, we show how we fit the multi-Gaussian emission line features with the help of {\sc ppxf} software. In \autoref{sec:ppxf}, we will first briefly introduce {\sc ppxf}. In \autoref{sec:kin_constr},  we describe the kinematic constraints on broad and narrow emission lines. In \autoref{sec:stellar_kin}, we acquire new stellar kinematics in a few galaxies where the DAP stellar continuum are not accurate due to the pollution of the broad lines. 

\subsection{ The {\sc ppxf} software}
\label{sec:ppxf}

Our emission line model is based on the latest version\footnote{Python \textsc{ppxf} package v8.2 from \url{https://pypi.org/project/ppxf/}} of {\sc ppxf} \citep{ppxf2004,pPXF,Cappellari2022}. This penalized pixel-fitting ({\sc ppxf}) software pioneered a robust pixel-fitting method, particularly optimized for extracting the kinematics of the stars and gas in galaxies from integral-field spectroscopic (IFS) data. 

When modeling gas emission lines, {\sc ppxf} fits the gas emission lines together with the stellar kinematics. Multiple Gaussian emission lines can be fitted by passing to the program an arbitrary number of Gaussian emission line templates. The kinematics of lines can be tied to each other, or they can be fitted independently. And the relative fluxes of emission line doublets can be fixed or fitted independently. Detailed constraints applied in this paper are described in the following section. 

\subsection{kinematic constraints for the gas emission lines}
\label{sec:kin_constr}

For the 1652 1$\sigma$ outliers identified in \autoref{sec:sample}, we opted not to make the multi-Gaussian fitting too complex and risk losing physical significance. Therefore, we assumed three multi-Gaussian models, each of which corresponds to different dynamic processes and is limited by different kinematic constraints. Specifically, the models contained either (1) broad Balmer lines, (2) broad Balmer and broad [OIII] lines, or (3) double-peaked narrow lines. We then selected the best fit model for each outlier and obtained the corresponding line properties.

Depending on the presence of broad [OIII] doublets,  \autoref{tab:emiconstrain1}, \autoref{tab:emiconstrain2}, and \autoref{tab:emiconstrain3} demonstrates the gas kinetic constraints on emission lines during our fitting. Lines included in gas kinematic component are given in the Tables and are fitted with the same velocity and dispersion.  \autoref{tab:emiconstrain1} shows the kinematic constraints in galaxies with broad balmer emission lines including broad $\rm H\delta$, $\rm H\gamma$, $\rm H\beta$ and $\rm H\alpha$. We use three different sets of emission line templates: (i) Narrow Balmer, (ii) Broad Balmer and (iii) Narrow forbidden. We use the \texttt{`constr\_kinem'} keyword in \textsc{ppxf} to set linear constraints on the emission line kinematics. We limit the velocity difference of broad and narrow Balmer lines to be less than broad line velocity dispersion to ensure the emission lines are at the corresponding wavelength. And the velocity dispersion of broad lines should be at least 600km/s larger than that of narrow lines to separate them apart. 

\begin{table}
  \centering
  \caption{Kinematic constraints in Galaxies with broad Balmer emission lines.}
  \begin{tabular}{|c|c|c|}
    \hline
    \hline
    Template name & Lines included & Kinematic parameters \\
    \hline
    Narrow Balmer & $\rm H\delta$, \;$\rm H\gamma$,\;$\rm H\beta$,\; $\rm H\alpha$  & $\rm V=V_{1}$,\; $\rm \sigma = \sigma_{1}$\\
     Broad Balmer & $\rm H\delta$, \;$\rm H\gamma$, \;$\rm H\beta$, \;$\rm H\alpha$  & $\rm V=V_{2}$, $\rm \sigma = \sigma_{2}$\\
     Narrow forbidden & $\rm [OIII]$, \;$\rm [NII]$, \;$\rm [SII]$  & $\rm V=V_{3}$, $\rm \sigma = \sigma_{3}$\\
    \hline
  \end{tabular}
  \text{constraints: $|V_{1}-V_{2}|<\sigma_{2}$, $\sigma_{2}>\sigma_{1}+600$ km~s$^{-1}$}
  \label{tab:emiconstrain1}
\end{table}
 \autoref{tab:emiconstrain2} shows the kinematic constraints in galaxies with broad balmer and broad [OIII] emission. Here we add an extra broad [OIII] template. While still using the dynamic restrictions on the broad Balmer lines, we limit the velocity difference of broad and narrow [OIII] lines to be less than broad line velocity dispersion. And the velocity dispersion of broad [OIII] should be at least 300 km/s larger than that of narrow forbidden lines. 
\begin{table}
  \centering
  \caption{Kinematic constraints in Galaxies with broad [OIII] and broad Balmer emission lines}
  \begin{tabular}{|c|c|c|}
    \hline
    \hline
    Template name & Lines included & Kinematic parameters \\
    \hline
    Narrow Balmer & $\rm H\delta$, $\rm H\gamma$, $\rm H\beta$, $\rm H\alpha$  & $\rm V=V_{1}$, $\rm \sigma = \sigma_{1}$\\
     Broad Balmer & $\rm H\delta$, $\rm H\gamma$, $\rm H\beta$, $\rm H\alpha$  & $\rm V=V_{2}$, $\rm \sigma = \sigma_{2}$\\
     Narrow forbidden & $\rm [OIII]$, $\rm [NII]$, $\rm [SII]$  & $\rm V=V_{3}$, $\rm \sigma = \sigma_{3}$\\
     Broad [OIII] & [OIII]4960, [OIII]5007  & $\rm V=V_{4}$, $\rm \sigma = \sigma_{4}$\\
    \hline
  \end{tabular}
  \text{constraints:$|V_{1}-V_{2}|<\sigma_{2}$,\, $\sigma_{2}>\sigma_{1}+600$ km~s$^{-1}$,\,
  $|V_{3}-V_{4}|<\sigma_{2}$}
  \label{tab:emiconstrain2}
\end{table}
 \autoref{tab:emiconstrain3} shows the kinematic constraints in galaxies with double narrow lines emission. For Balmer lines and forbidden lines, two sets of templates are given to each set of lines. We limit the velocity difference of the same lines in different templates to be larger than three times the velocity dispersion to ensure they are separated apart. 
\begin{table}
  \centering
  \caption{Kinematic constraints in Galaxies with double-peaked narrow lines.}
  \begin{tabular}{|c|c|c|}
    \hline
    \hline
    Template name & Lines include & Kinematic parameters \\
    \hline
    Narrow Balmer 1 & $\rm H\delta$, $\rm H\gamma$, $\rm H\beta$, $\rm H\alpha$  & $\rm V=V_{1}$, $\rm \sigma = \sigma_{1}$\\
    Narrow Balmer 2 & $\rm H\delta$, $\rm H\gamma$, $\rm H\beta$, $\rm H\alpha$  & $\rm V=V_{1}$, $\rm \sigma = \sigma_{2}$\\
     Narrow forbidden 1 & $\rm [OIII]$, $\rm [NII]$, $\rm [SII]$  & $\rm V=V_{3}$, $\rm \sigma = \sigma_{3}$\\
     Narrow forbidden 2 & $\rm [OIII]$, $\rm [NII]$, $\rm [SII]$  & $\rm V=V_{4}$, $\rm \sigma = \sigma_{4}$\\
    \hline
  \end{tabular}
   \text{constraints: $V_{1}-V_{2}>72$ km~s$^{-1}$,\, $V_{3}-V_{4}>72$ km~s$^{-1}$. }
  \label{tab:emiconstrain3}
\end{table}

The possible combination of double-peaked narrow line and broad line is not included in  \autoref{tab:emiconstrain1}, \autoref{tab:emiconstrain2}, and \autoref{tab:emiconstrain3}. We find only one galaxy with such rare emission line feature. The galaxy is fitted manually and we will discuss its dynamic features in detail in \autoref{sec:combination of lines}.

For the majority of galaxies we adopt the stellar continuum provided by DAP. So we do not fit polynomials with {\sc ppxf}. In a few cases, the width of the broad line affects the stellar continuum fitting of the DAP. These cases are discussed in \autoref{sec:stellar_kin}. \autoref{fig:egbroadha}, \autoref{fig:egbroadOIII} and \autoref{fig:egdualnarrow}, show examples of broad $\rm H\alpha$ or dual narrow line fitting. We did not fit for attenuation here. And considering the time taken, we did not perform a global optimization with {\sc ppxf} keyword \texttt{`globel\_search'} on every galaxy. However, we compared the convergence in our fits for broad line galaxies with the global optimized fits using \texttt{`globel\_search'} and find little difference between them. Detailed emission line properties and galaxy information can be found in appendix~\ref{Appendix A}. 

\subsection{Stellar kinematic of galaxies with very broad emission}
\label{sec:stellar_kin}

The presence of broad emission lines also affects the stellar
kinematics produced by DAP, which masks the emission lines according to a specific width of 800 km/s and determines the stellar kinematics using the rest of the spectrum \citep{DAP}. The width is insufficient for broad emission lines, especially broad Balmer lines with typical with of $\sim$1000 km/s. Insufficient masking of the emission line area leads to inaccurate kinetic properties in some spaxels. 

\begin{figure*}
	\includegraphics[width=\textwidth]{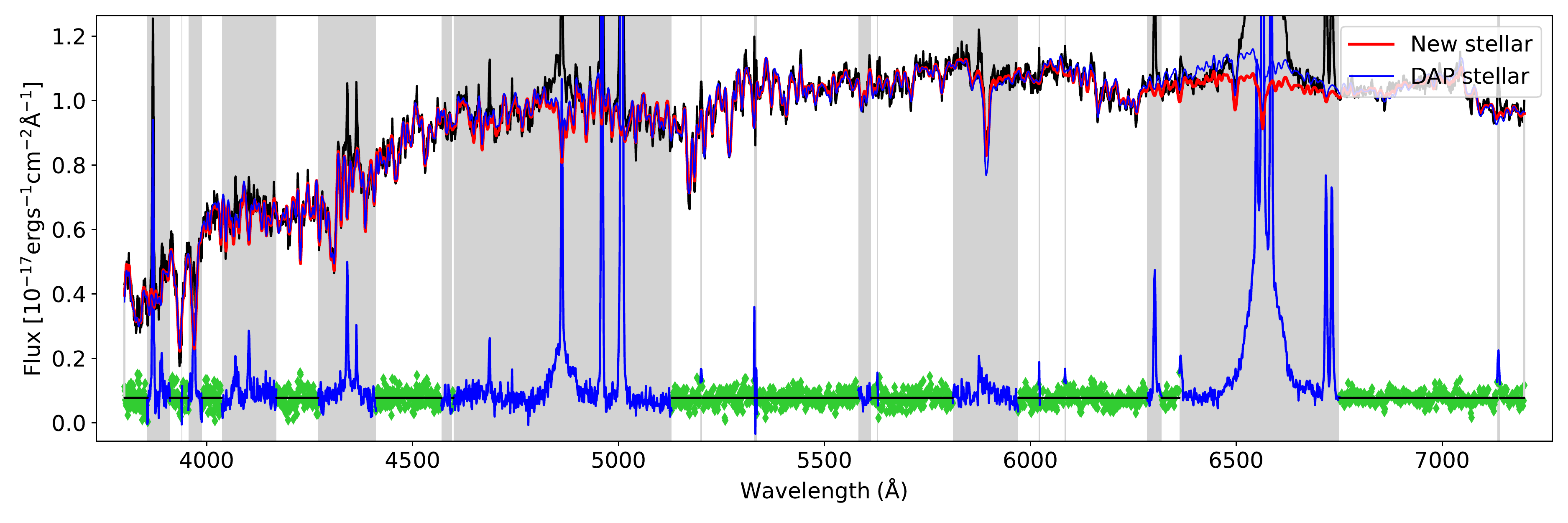}
    \caption{Example fit to the stellar continuum with \textsc{ppxf} to extract the stellar kinematics (MaNGA-ID: 1-495383, 1-53093 ,1-576315 from top to bottom). The shaded areas are those that are blocked due to the influence of the broad line. A typical width of broad Balmer lines in these severely affected galaxies is broad $\rm \sigma_{H\alpha}\sim$ 1500 km/s. We masked a 3$\rm \sigma$ range of $\pm$4500 km/s around $\rm H\alpha$ and a 2$\rm \sigma$ range of $\pm$3000 km/s around $\rm H\gamma$ and $\rm H\delta$. Finally, the $\rm H\beta$ and [OIII] region is masked completely within $\pm$6000 km/s around $\rm H\beta$ because this region is heavily contaminated. The black lines are the actual spectra. The blue lines to the top are the stellar continuum from DAP. The red lines are our new stellar continuum. The unnatural protrusions in $\rm H\alpha$ area disappear. The green diamond point shows the residuals which exclude the masked emission lines.}
    \label{fig:correct-8561-3704}
\end{figure*}

After examining the broad-line galaxies, those galaxies where $\rm H\alpha$ lines width are around 1000 km/s are not much affected
because nearby [NII] doublets help mask the entire $\rm H\alpha$ region in the spectrum. We acquire the new stellar continuum of the galaxies where broad $\rm \sigma_{H\alpha}$ > 1200 km/s by masking broad-line regions in spectra. Three examples are shown in  \autoref{fig:correct-8561-3704}. 

\begin{figure*}
	\includegraphics[width=\textwidth]{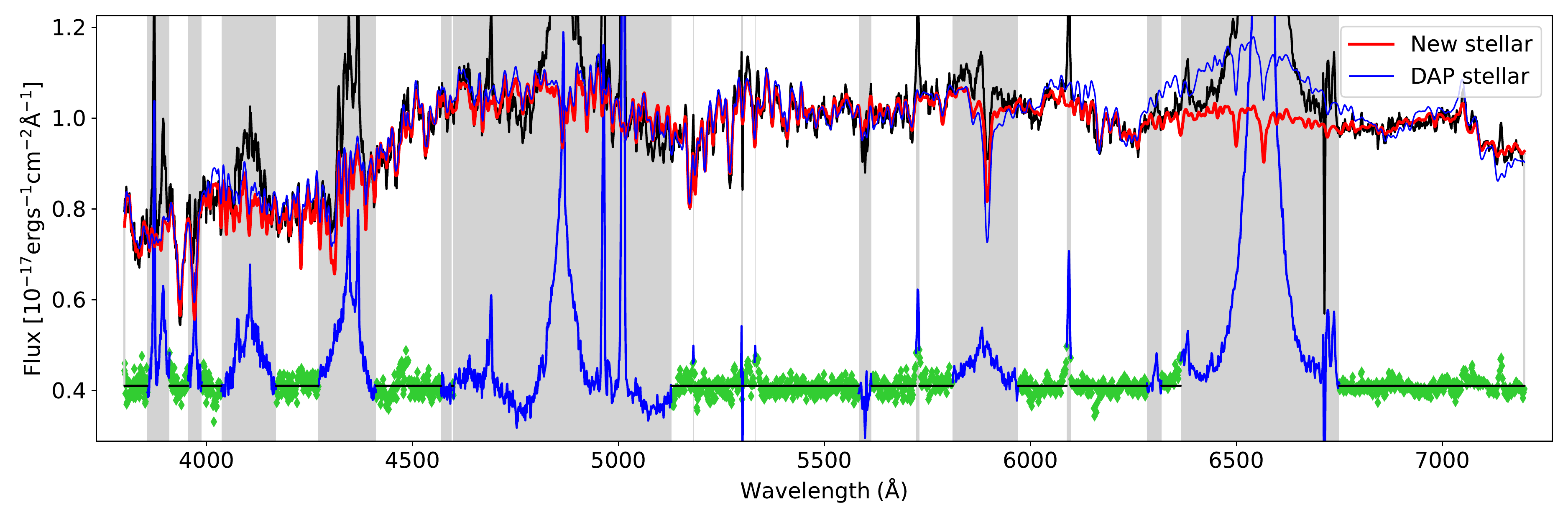}
\end{figure*}
\begin{figure*}
	\includegraphics[width=\textwidth]{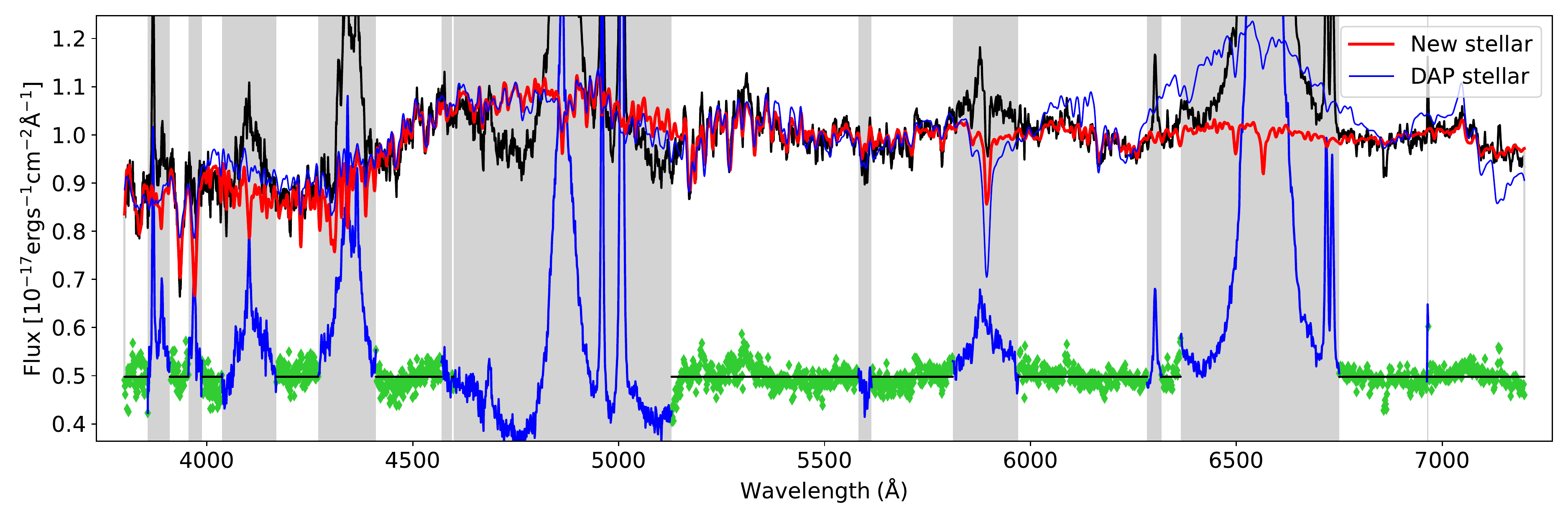}
\end{figure*}

\begin{figure*}
	\includegraphics[width=\textwidth]{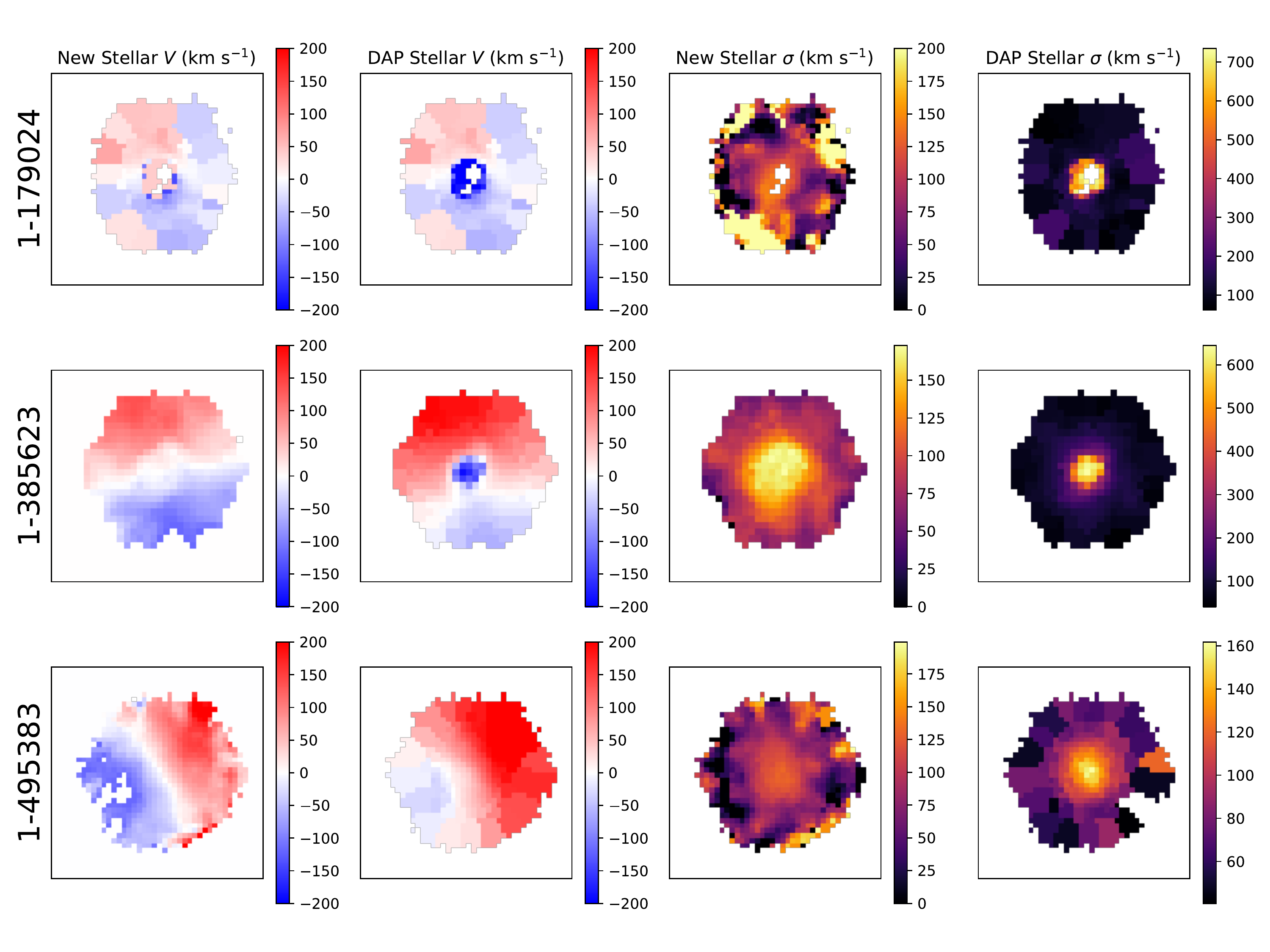}
    \caption{Comparison of stellar kinematics maps between our fits and DAP. The MaNGA-IDs of the three example galaxies are shown to the left of each row. Our fits have eliminated the apparently unphysical negative velocities in the broad line region at the center of the galaxy (As can be seen from the first two rows) and corrected the asymmetry in velocity maps due to wrong velocity in center spaxels influenced by AGNs. (As can be seen from the last row)}
    \label{fig:correct-sigma}
\end{figure*}

In the process of fitting the new galaxy continuum, we also obtain new values for stellar kinematic properties. Only in some of these galaxies, they are very different from DAP stellar kinematic properties.  \autoref{fig:correct-sigma} gives an example.

\subsection{Fitting Result}
\label{sec:fitresult}
We create a catalogue containing multi-Gaussian line properties of MaNGA galaxies derived from {\sc ppxf} method.  \autoref{tab:examplecatalog} reports how we present our results. 

\autoref{tab:examplecatalog} and \autoref{tab:examplecatalog2} gives 10 example galaxies. The luminosity ratio is measured using the apparent luminosities of two narrow $\rm H_\alpha$ components. Using $L_{\rm H\alpha}$ with larger velocity as numerator divide by $L_{\rm H\alpha}$ with smaller velocity as denominator. And $\Delta V$ is the absolute velocity difference measure from the center of the velocity. An entire catalogue can be found in appendix~\ref{Appendix A}. \HL{Comparing our broad line galaxies with the previous samples of type-1 AGNs in \cite{Sanchez_2022}. We recovered broad lines for 104 of the 119 type-1 AGN candidates in \cite{Sanchez_2022}, and we discovered 35 new galaxies with broad emission lines. For the rest 15 examples in \cite{Sanchez_2022}, we believe that these galaxies only have narrow lines with relatively larger velocity dispersion ($\sim$250km/s). We did not find broad Balmer emission lines. They are actually type-2 AGN candidates misclassified as type-1.}

\begin{table*}
  \centering
  \caption{Example catalogue of 5 galaxies with broad line features in MaNGA. A machine-readable version of the full table is available (see Supporting Information).}
  \begin{tabular}{llllllll}
    \hline
    MaNGA-ID & RA & Dec &$\lg L^{\rm broad}_{H\alpha}$ & $\lg\sigma^{\rm broad}_{H\alpha}$ & $\lg\sigma_{\rm e}$ & Extra templates \\
    & (deg) & (deg) & ( $10^{40}$erg s$^{-1}$) & (km s$^{-1}$) & (km s$^{-1}$) &\\
    \hline
    1-382273 & 118.634242 & 16.809729 & 12.91  & 3.403 & 2.137 & Broad H$\alpha$            \\
    1-547191 & 258.118529 & 35.884086 & 14.71   & 3.251 & 1.954 & Broad Balmer               \\
    1-300461 & 137.160233 & 32.592953 & 6.31   & 3.098 & 2.114 & Broad H$\alpha$            \\
    1-587259 & 160.228870  & 5.991890   & 12.61   & 2.951 & 1.964 & Broad H$\alpha$,{[}OIII{]} \\
    1-150947 & 183.263995 & 51.648597 & 47.03  & 2.958 & 2.332 & Broad H$\alpha$,{[}OIII{]} \\
    \hline
  \end{tabular}
  \begin{tablenotes}
    \item Note: Column (1) is the MaNGA identification ID. Column (2-3) is the position of the galaxy shown in celestial coordinates. We use the positions to acquire optical and radio images. Column (4) is the apparent luminosity of the broad $\rm H\alpha$ component. Column (5) is the flux-weighted velocity dispersion of the broad $\rm H\alpha$ component. Column (6) is the stellar velocity dispersion within 1Re corrected by masking the broad line region in \autoref{sec:kin_constr}. Column (7) shows the extra templates added when fitting each galaxy. The label \texttt{`Broad H$\alpha$'} means in the broad Balmer set of templates the broad $\rm H\delta$, $\rm H\gamma$ and $\rm H\beta$ components are all zero or below the noise level. Only broad $\rm H\alpha$ is observable. 
  \end{tablenotes}
  \label{tab:examplecatalog}
\end{table*}

\begin{table*}
  \centering
  \caption{Example catalogue of 5 galaxies with double narrow line features in MaNGA. A machine-readable version of the full table is available (see Supporting Information).}
  \begin{tabular}{llllllll}
    \hline
    MaNGA-ID & Ra & Dec &Luminosity Ratio&  $\lg \Delta V$ (km/s) & $\lg \sigma_{e}$ (km/s) & Extra templates \\
    \hline
    1-114955 & 332.602090  & 11.713077 & 1.91   & 2.444 & 2.433 & Dual narrow                \\
    1-596678 & 332.892838 & 11.795929 & 0.53   & 2.358 & 2.130  & Dual narrow                \\
    1-115365 & 333.483204 & 13.755396 & 1.10    & 2.391 & 1.982 & Dual narrow                \\
    1-42007  & 33.626933  & 13.257206 & 1.61   & 2.367 & 2.057 & Dual narrow                \\
    1-41752  & 31.953063  & 13.609595 & 3.01   & 2.330  & 2.233 & Dual narrow                \\
    \hline
  \end{tabular}
  \begin{tablenotes}
  \item Note: Column (1) is the MaNGA identification ID. Column (2-3) is the position of the galaxy shown in celestial coordinates. We use the positions to acquire optical and radio images. Column (4) is the apparent luminosity ratio measured through the double $\rm H\alpha$ components. Column (5) is the velocity difference between the two narrow components. Column (6) is the stellar velocity dispersion within 1Re. Column (7) shows the extra templates added when fitting each galaxy. The label \texttt{`Dual narrow'} means adding an extra set of narrow lines with kinematic constrains from \autoref{tab:emiconstrain3}. 
  \end{tablenotes}
  \label{tab:examplecatalog2}
\end{table*}

%% file: discussion.tex
In this section, we will discuss how galaxies with broad emission lines are characterized based on the emission line properties given in this paper. In \autoref{sec:BPT diagrams}, we will discuss the classification of broad-line galaxies on BPT diagrams. In \autoref{sec:Dynamic properties} and \autoref{sec:M-s relation}, we want to study the AGN-galaxy coevolution through the stellar population of broad-line region host galaxies and the relation between broad lines' properties and the host galaxies' dynamical properties. In \autoref{sec:dual AGN}, we show examples of how the spatial distribution of broad emission lines can help us find AGN mergers.

\subsection{BPT diagrams}
\label{sec:BPT diagrams}

For galaxies with broad emission lines, we not only fit the broad lines but also correct the misfit of the narrow line properties in DAP caused by the broad lines. The relative strengths of different narrow lines provide sensitive diagnostics of the properties of the ionized gas and the astrophysical processes in their host galaxies.

MaNGA’s observed wavelength range (3600 - 10300 \AA) covers all the strong nebular lines used in the standard BPT diagram \citep{BPTorigin} in MaNGA redshift range (0.02<z<0.15) . Based on MaNGA data, \cite{3DBPT} finds that the gas phase velocity dispersion correlates strongly with traditional optical emission line ratios such as [SII]/$\rm H\alpha$, [NII]/$\rm H\alpha$, and [OIII]/$\rm H\beta$. Henceforth we adopt the definition in \cite{3DBPT} of some most widely used ratios such as $\rm \log ([NII] \lambda6585/H\alpha$) (hereafter `N2'), $\rm \log ([SII] \lambda6718+6732/H\alpha$) (hereafter `S2'), and $\rm \log ([OIII] \lambda5008/H\alpha$) (hereafter `R3'). The corresponding relation that separates the cold-disk sequence from the warm-disk sequence takes the form :
\begin{equation}
N2 = \sum_{i=0}^2 \sum_{j=0}^2 C_{ij} S2^i R3^j ,
\label{BPTclass}
\end{equation}
where $C_{ij}$s are the coefficients shown in  \autoref{tab:C}.
\begin{table}
  \centering
  \caption{Coefficients of the polynomial surface in \autoref{BPTclass}. This is Table~2 of \citet{3DBPT}.}
  \begin{tabular}{|c|c|c|c|}
    \hline
    $C_{ij}$ & j = 0 & 1 & 2 \\
    \hline
    i=0 & $-$0.7362 & $-$0.6464 & $-$1.7567 \\
    1 & $-$1.2338 & $-$2.0170 & 1.3520 \\
    2 & $-$0.3036 & 0.4533 & 0.3177 \\
    \hline
    \end{tabular}
  \label{tab:C}
\end{table}

An edge-on projection of the interface is shown under coordinates $P1$, $P2$ where \citep[eq.~12-13]{3DBPT}.
\begin{equation}
P1=0.77\, N2 + 0.54\, S2 + 0.33\, R3 .
\end{equation}
\vspace{-0.5cm}
\begin{equation}
P2=-0.57\, N2 + 0.82\, S2 . 
\end{equation}

In  \autoref{fig:BPTexample}, we illustrate the
types of ionization structures observed in MaNGA galaxies and our classification scheme. The galaxy in  \autoref{fig:BPTexample}a (MaNGA-ID: 1-523004) is a typical star-forming galaxy with broad $\rm H\alpha$ emission. Star-forming spaxels dominate the broad line region around its galactic nucleus. 

The galaxies in  \autoref{fig:BPTexample}b and \autoref{fig:BPTexample}c are Classified as AGN with broad $\rm H\alpha$ emission. AGN spaxels dominate the broad line region. Then we put spaxels that are classified as AGN in the S2 vs. R3 diagram and separate LINERs from Seyfert AGNs. 

\subsection{Stellar population of AGNs}
\label{sec:Dynamic properties}
\begin{figure}
	\includegraphics[width=0.5\textwidth]{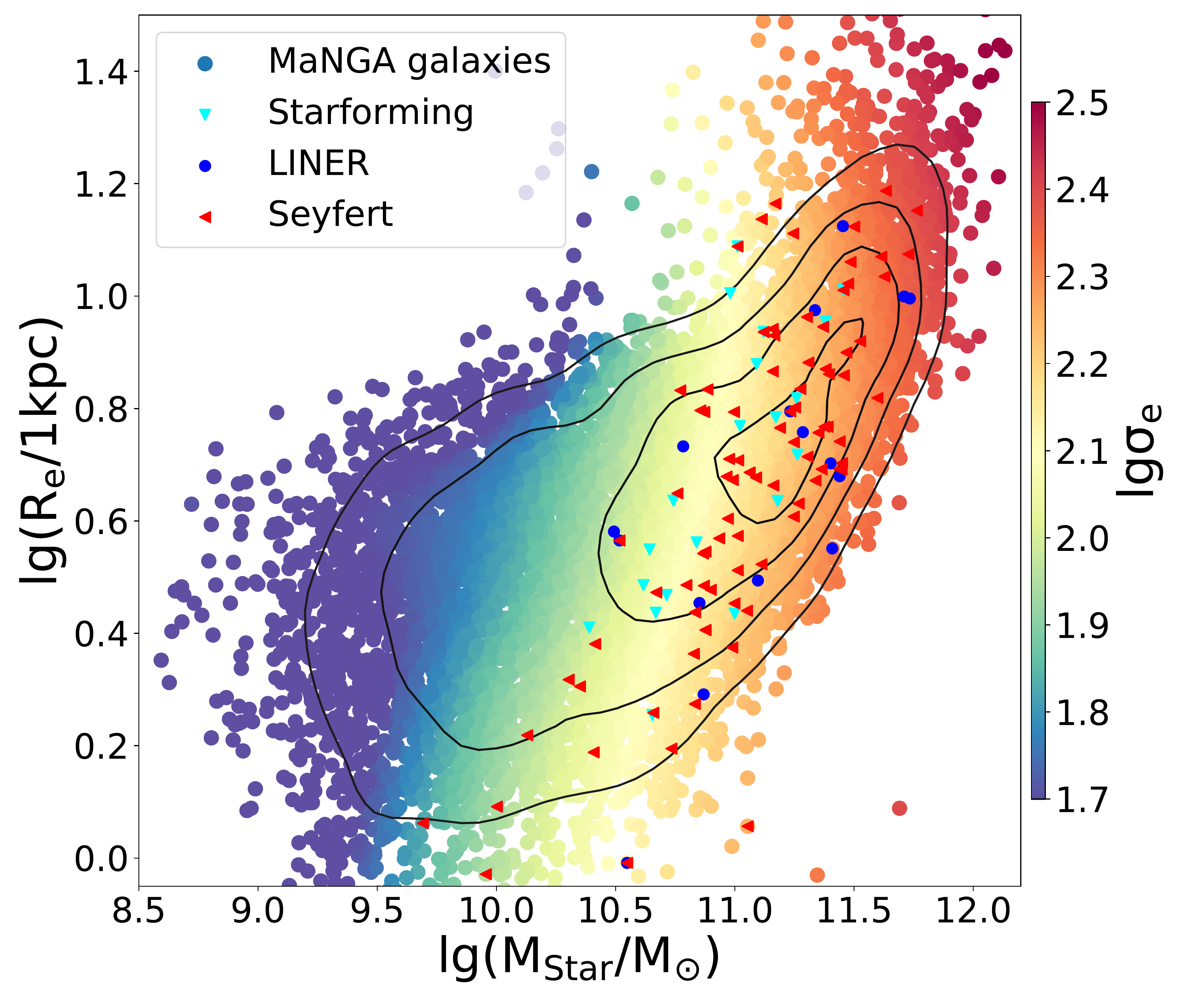}
    \caption{The distribution of mass vs. effective radius of broad $\rm H\alpha$ galaxies. The dots coloured by $\rm \sigma_e$ from DAP in the background represent the distribution of all other MANGA galaxies. Galaxies classified into different categories are represented using different colours. The contour lines are a kernel density estimate of the galaxies number density, computed with the Scipy \citep{Scipy2020} function \href{https://docs.scipy.org/doc/scipy/reference/generated/scipy.stats.gaussian_kde.html}{\texttt{scipy.stats.gaussian\_kde}}. The velocity dispersion was smoothed using the \textsc{LOESS}  procedure by \citet{Cappellari2013p20}, which implements the algorithm by \citet{cleveland1988locally}.}
    \label{fig:M-size}
\end{figure}

\begin{figure}
	\includegraphics[width=0.5\textwidth]{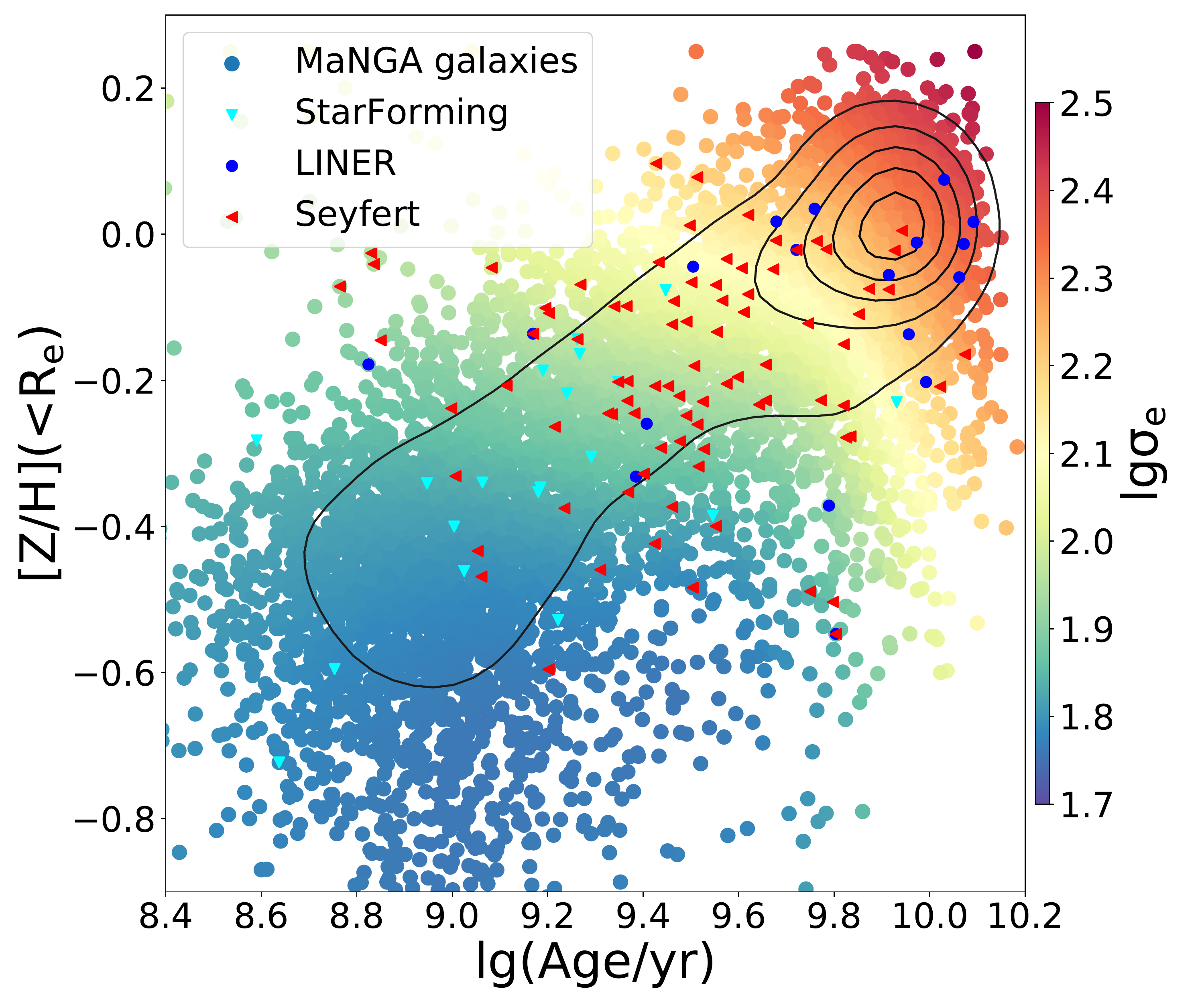}
    \caption{The distribution of age vs. metallicity of broad $\rm H\alpha$ galaxies. The colours and symbols are described in  \autoref{fig:M-size}.}
    \label{fig:metal}
\end{figure}

One popular method to study the AGN-galaxy coevolution is via examining the stellar population of AGN host galaxies. \autoref{fig:M-size} and \autoref{fig:metal} show the distribution of mass, effective radius, age, and metallicity of broad $\rm H\alpha$ galaxies classified into different categories by BPT diagram. The coloured dots in the background represent the distribution of all other MANGA galaxies under the same set of physical parameters. The stellar population properties used in this work is from \citet{LuShengdong2023}, which applies the {\sc ppxf} \citep{Cappellari2022} on the full sample in MaNGA project and derives their \HL{both global and} spatially resolved stellar population properties (including stellar mass, metallicity, and age). \HL{The stellar masses used here are calculated within the elliptical half-light isophotes, taken from the catalogue of \citet{LuShengdong2023} under the keyword \texttt{`Mstar\_Re'}.} The ages and metallicities are the SDSS $r-$band \citep{Stoughton_et_al_2002} luminosity-weighted average value within the half-light isophote under keyword \texttt{`LW\_Age\_Re'} and \texttt{`LW\_Metal\_Re'}. The stellar velocity dispersions within 1Re come from DAPall file \citep{DAP} and the effective radius (Re) are half-light radius provided by the NSA catalogue. The NSA catalogue is an extension of the NASA-Sloan Atlas (NSA)\footnote{M. Blanton; \url{www.nsatlas.org}} toward higher redshift (z $\leq$ 0.15) and includes an elliptical Petrosian analysis of the surface-brightness distributions \citep{NSA_catalog}.

According to \autoref{fig:M-size}, AGN and star-forming triggered broad-line galaxies are scattered around the middle of the image. The host galaxies of LINER tend to have larger $\rm \sigma_e$ and more than half of them lie in the region of the elliptical galaxies to the upper right of the figure. This is consistent with the morphology observed from the images. 

\autoref{fig:metal} shows that the star-forming galaxies are relatively young and LINER galaxies are old. The old ages of LINERS is consistent with previous studies who concluded that LINER emission is due to ionization by evolved stars \citep{Sarzi2010,Yan2012,2016MNRAS.461.3111B,Seyfert_2021}. Compared with the number density of MANGA galaxies, the distribution of AGN host galaxies are not concentrated in the upper-right region. Instead, they show a slightly scattered distribution in the middle. The AGN host galaxies occupy a large range of ages from old to young galaxies and they preferentially lie around intermediate ages. The finding corroborate previous results in \cite{AGN_age2} and \cite{AGN_age}. \HL{The relatively young age of AGN host galaxies hints at the presence of a significant younger stellar population \citep{Age_2003,Age_2004} that is consistent with the result through color analyses \cite[eg.,][]{2014MNRAS.440..889S,color_2016}.}

\begin{figure}
	\includegraphics[width=0.5\textwidth]{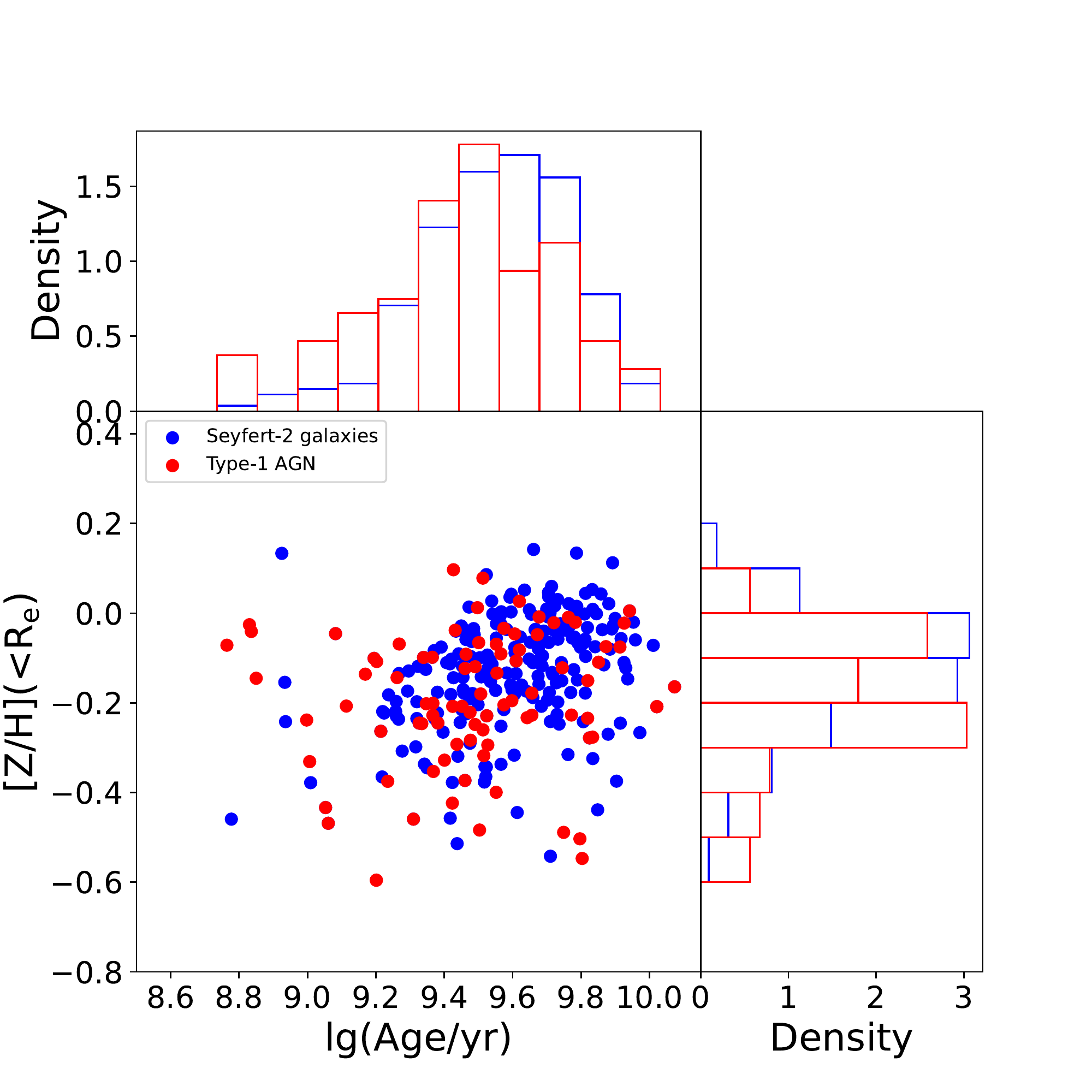}
    \caption{The distribution of age vs. metallicity of broad $\rm H\alpha$ type-1 AGNs and narrow line seyfert-2 galaxies. }
    \label{fig:1v2}
\end{figure}

\HL{\cite{Seyfert_2021} described in detail the classification criterion for different sources of ionization using traditional BPT diagrams as well as the equivalent width of $\rm H\alpha$ and the spatial distribution of line ratios. An AGN ionized region is characterized by its emission line ratios well above the \cite{K01} demarcation lines and EW($\rm H\alpha$) larger than 3 \AA. They present a decrease of the considered line ratios with respect to the central values in the galaxy and show a steep decline in the flux intensity \citep{Seyfert_2020,Seyfert_2021}. In most of our broad line galaxies, broad line regions largely coincide with AGN ionized regions. To further understand the role of the broad line in AGN ionized region, }\autoref{fig:1v2} shows the distribution of broad-line type-1 AGNs and narrow-line seyfert-2 host galaxies on the age-metallicity diagram. \HL{Due to the fact that AGN ionized region is mostly concentrated in the central regions \cite[eg.][]{Husemann_2010,Husemann_2014,Seyfert_2020}} the seyfert-2 host galaxies are classified by applying the standard BPT diagram on the centre spaxel and obtaining the \texttt{`seyfert'} classifications without broad lines \HL{as a first order approximation}. Our work indicates that the host galaxy of type-1 AGNs are slightly younger than seyfert-2 galaxies. Their metallicity distribution follows the same trend. \cite{AGN_age} recently explored the properties of the host galaxies of X-ray selected AGN in the COSMOS field using the Chandra Legacy sample and the LEGA-C survey VLT optical spectra. They measured the ages of different types of AGNs through $\rm D_n$4000. Our finding is consistent with their result.

\subsection{Testing the $M_{\rm BH}-\sigma_{\rm e}$ relation in type-1 AGNs}
\label{sec:M-s relation}

Another way to study the AGN-galaxy coevolution is to connect broad lines' properties to the galaxies' dynamical properties. \cite{M-broad} suggested that the velocity dispersion and luminosity of broad lines have a strong correlation with black hole mass in active galactic nuclei. Meanwhile, the black hole mass is related to the stellar velocity dispersion within the effective radius ($\sigma_e$). Previously, the $M_{\rm BH}-\sigma_{\rm e}$ relation \citep{Gebhardt2000bh,Ferrarese2000} has been parameterized as a power-law function ($M_{\rm BH} \propto \sigma_{\rm e}^{\alpha}$) \citep[e.g.][]{Beifiori_2012,McConnell_2013,Kormendy_2013,M-S2015,M-sigmastar}, where $\alpha$ was found to be between 3 and 6. 

For type-1 AGN with broad H$\alpha$ lines, we used the prescription of \citet[][eq.6]{M-broad} : 
\begin{equation}
M_{\rm BH} = 2\times 10^6 \left(\frac{L_{\rm broad\,H\alpha}}{\rm 10^{42}erg\,s^{-1}}\right)^{0.55} \left(\frac{{\rm FWHM}(H\alpha)}{1000\; \rm km/s}\right)^{2.06},
\end{equation}
where ${\rm FWHM}(H\alpha)$ is  calculated from the flux weighted average of broad $\rm H\alpha$ velocity dispersion in different spaxels. Notice that ${\rm FWHM}(H\alpha) = \sigma\sqrt{4 \ln(4)}$. \autoref{sec:methods} provides the apparent luminosity of the broad $\rm H\alpha$ line. The attenuation-corrected $\rm H\alpha$ flux is obtained through \autoref{eq:dustcorrection}: 

\begin{equation}
L_{\rm broad\,H\alpha,corr} = L_{\rm broad\,H\alpha}10^{0.4A_{\rm H\alpha}},
\label{eq:dustcorrection}
\end{equation}
where $A_{\rm H\alpha} = 3.33\;E(B-V)$ according to \citet[eq 1,6]{BD}.
 
We measure E(B$-$V) from the Balmer decrement \citep[eq 4]{BD} : 
\begin{equation}
E(B-V) = 1.97\;\lg \left [\frac{\rm (H\alpha/H\beta)_{obs}}{2.86} \right ]. 
\label{eq:EBV}
\end{equation}

\begin{figure}
    \includegraphics[width=0.5\textwidth]{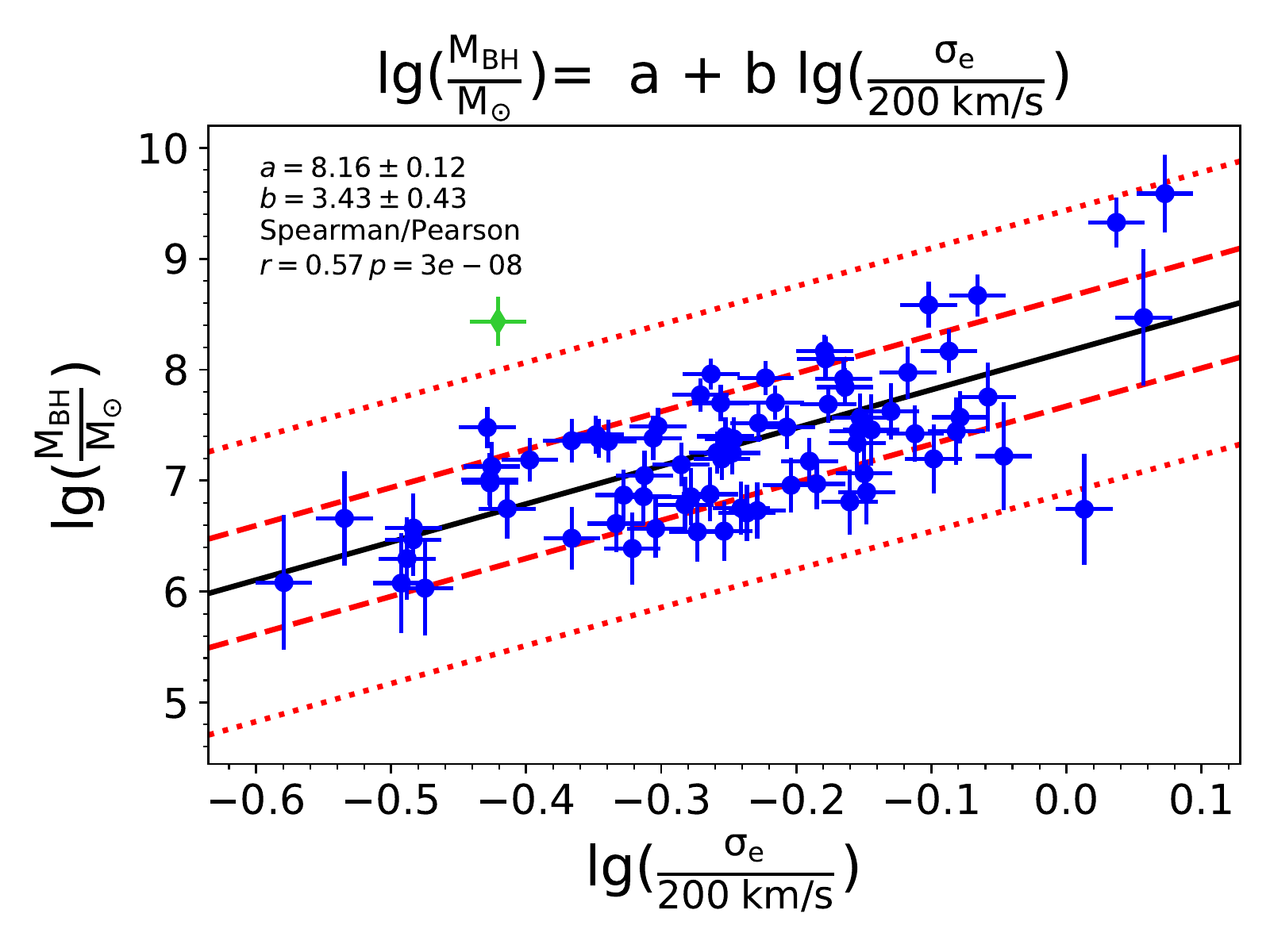}
    \caption{The $M_{\rm BH}-\sigma_{\rm e}$ relation in MaNGA type-1 AGNs. The blue dots are $\rm M_{BH}$ and $\rm \sigma_e$ of type-1 AGN with errorbar. The dashed red lines represent the 1$\sigma$ (68\%) and 2.6$\sigma$ (99\%) error range of the linear fit. The green dot is not included as it falls outside the range of fitting error. The fitting result is represented by the solid black line, while the fitting parameters, the Spearman-Pearson correlation value, and the probability of no correlation are displayed in the upper left corner of the figure.}
    \label{fig:sigmae}
\end{figure}

In \autoref{fig:sigmae}, we test the $\rm M_{BH}$-$\rm \sigma_e$ relation in our type-1 AGN catalogue. For the fit we used the \texttt{lts\_linefit} procedure\footnote{Available from \url{https://pypi.org/project/ltsfit/}} by \citet{Cappellari2013p15} which combines the Least Trimmed Squares robust technique by \citet{Rousseeuw2006} into a least-squares fitting algorithm. The procedure allows for errors in both variables and intrinsic scatter. We present our result in \autoref{eq:MSrelation} :  
\begin{equation}
\lg\left(\frac{M_{\rm BH}}{ M_{\rm \odot}}\right) = (8.16\pm 0.12) + (3.43\pm  0.43)\lg\left(\frac{\sigma_{\rm e}}{\rm 200km/s}\right).
\label{eq:MSrelation}
\end{equation}

Our result matches with the recent $M_{\rm BH}-\sigma_{\rm e}$ relation in \autoref{eq:MSrelationref} given by \citet[eq.14]{M-sigmastar} :
\begin{equation}
\lg\left(\frac{M_{\rm  BH}}{M_{\rm \odot}}\right) = (8.14\pm 0.20) + (3.38\pm  0.65)\lg\left(\frac{\sigma_{\rm e}}{\rm 200km/s}\right). 
\label{eq:MSrelationref}
\end{equation}

And the slope of our result is within the allowable range of \cite{M-S2015} ($3.97\pm0.56$).

\subsection{Discovery of dual AGNs}
\label{sec:dual AGN}
The multiple supermassive black holes are expected to exist inside many galaxies due to the previous merging events. They can be revealed by the detection of the coexistence of two active galactic nuclei in the merging system of galaxies \citep{dualAGN_nature}. \HL{Recently, \cite{merger_2023} studied the incidence of major mergers and their impact on the triggering of AGN activities using MaNGA DR15 sample.} Building a large sample of local dual AGNs amongst merging galaxies helps understand how the merging process of some galaxies excite galactic nuclei and convert them into AGN \citep{dual_2,dual_QSO}.

\begin{figure}
	\includegraphics[width=0.5\textwidth]{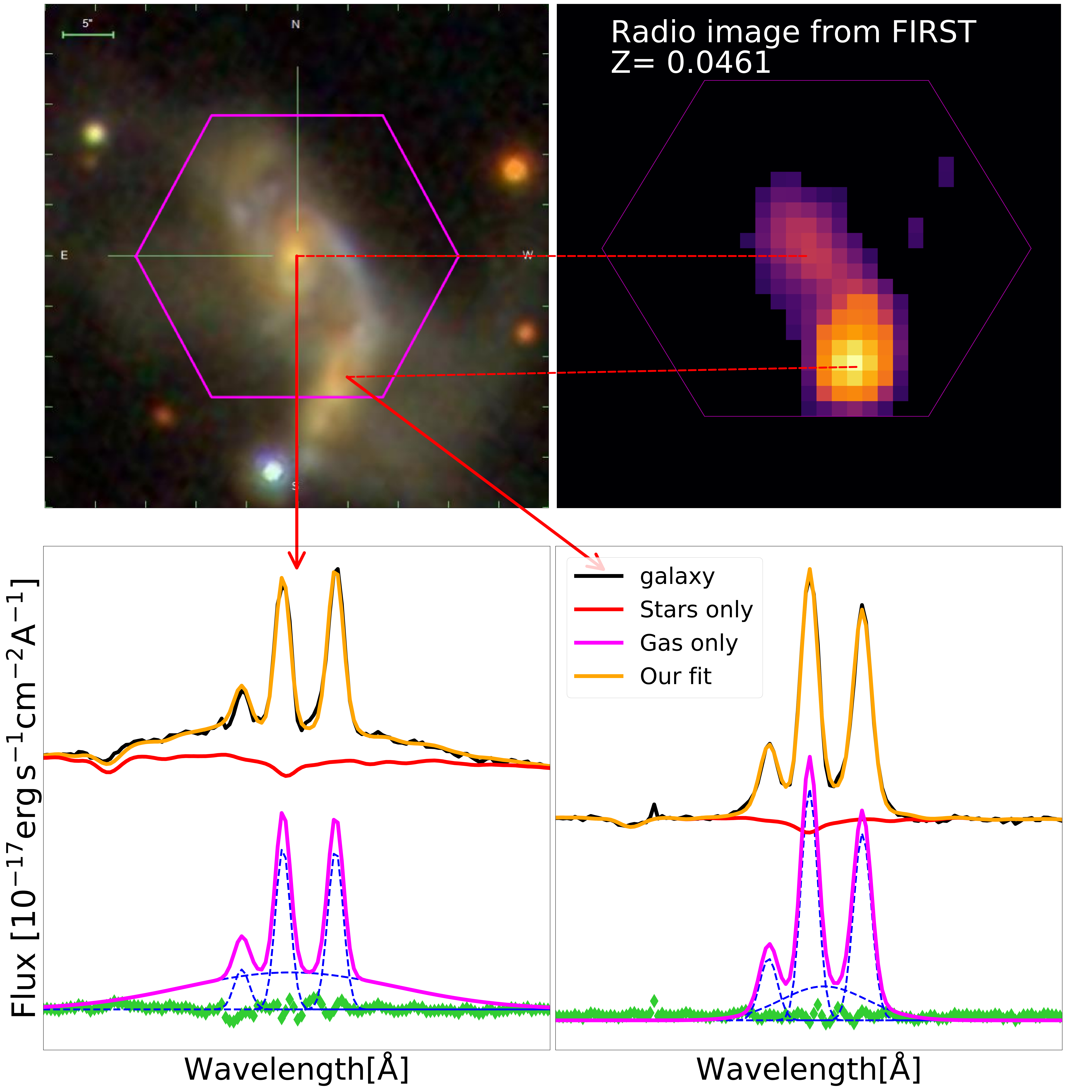}
    \caption{Image and emission line fitting result for dual AGN candidates (MaNGA-ID: 1-382273). The upper left panel is the galaxy image from MaNGA. The upper right panel shows the radio image from the FIRST radio survey \citep{Firstsurvey}. We can see a clear bimodal feature. Redshift and separation measured from flux centre is labeled on the radio image. The two images below show $\rm H\alpha$ fitting at the centres. The green dots represent the residual of multi-Gaussian fit. }
    \label{fig:dualAGN10218}
\end{figure}

\begin{figure}
	\includegraphics[width=0.5\textwidth]{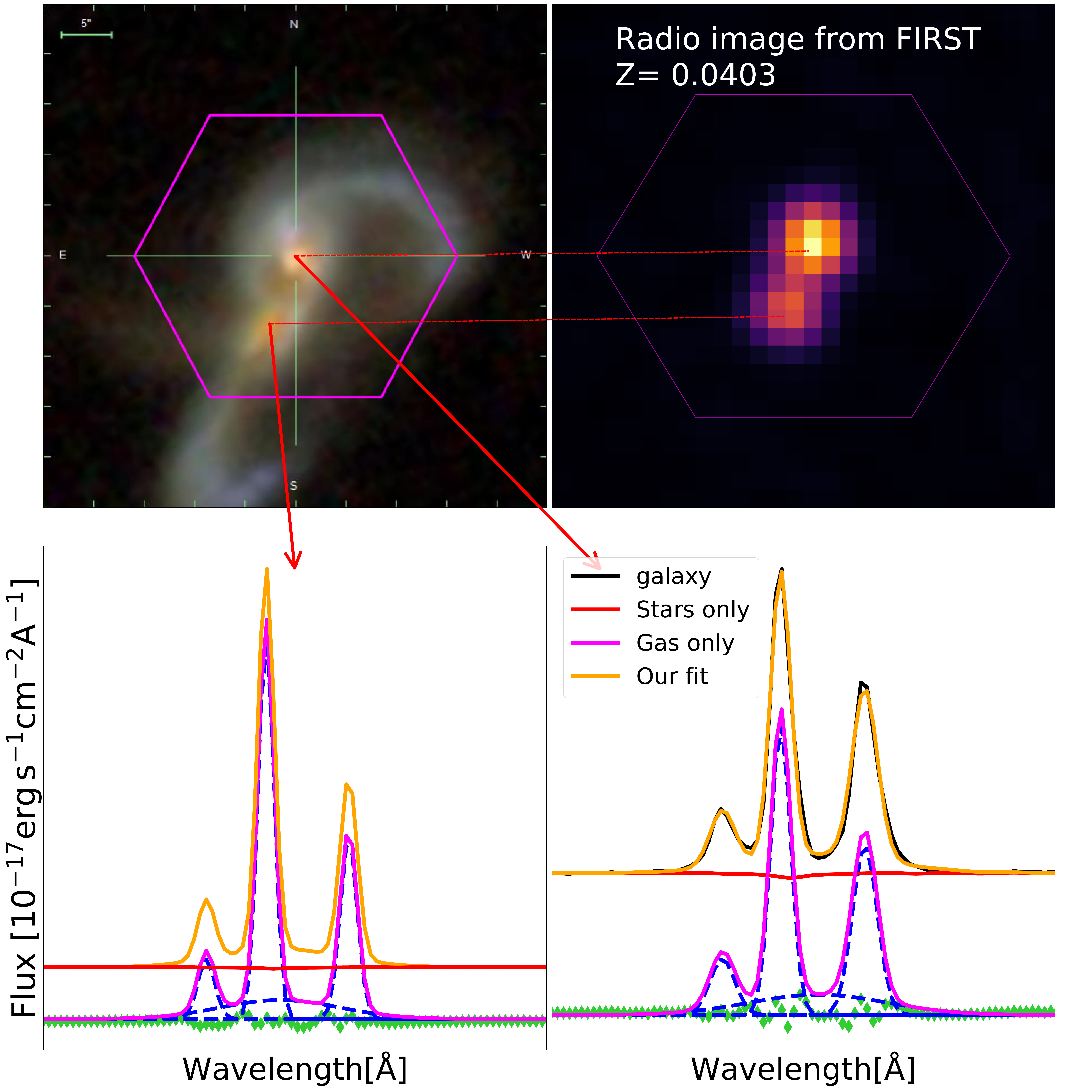}
    \caption{Image and emission line fitting result for dual AGN candidates (MaNGA-ID: 1-585513). The symbols are the same as in  \autoref{fig:dualAGN10218}. }
    \label{fig:dualAGN8250}
\end{figure}

\begin{figure}
	\includegraphics[width=0.5\textwidth]{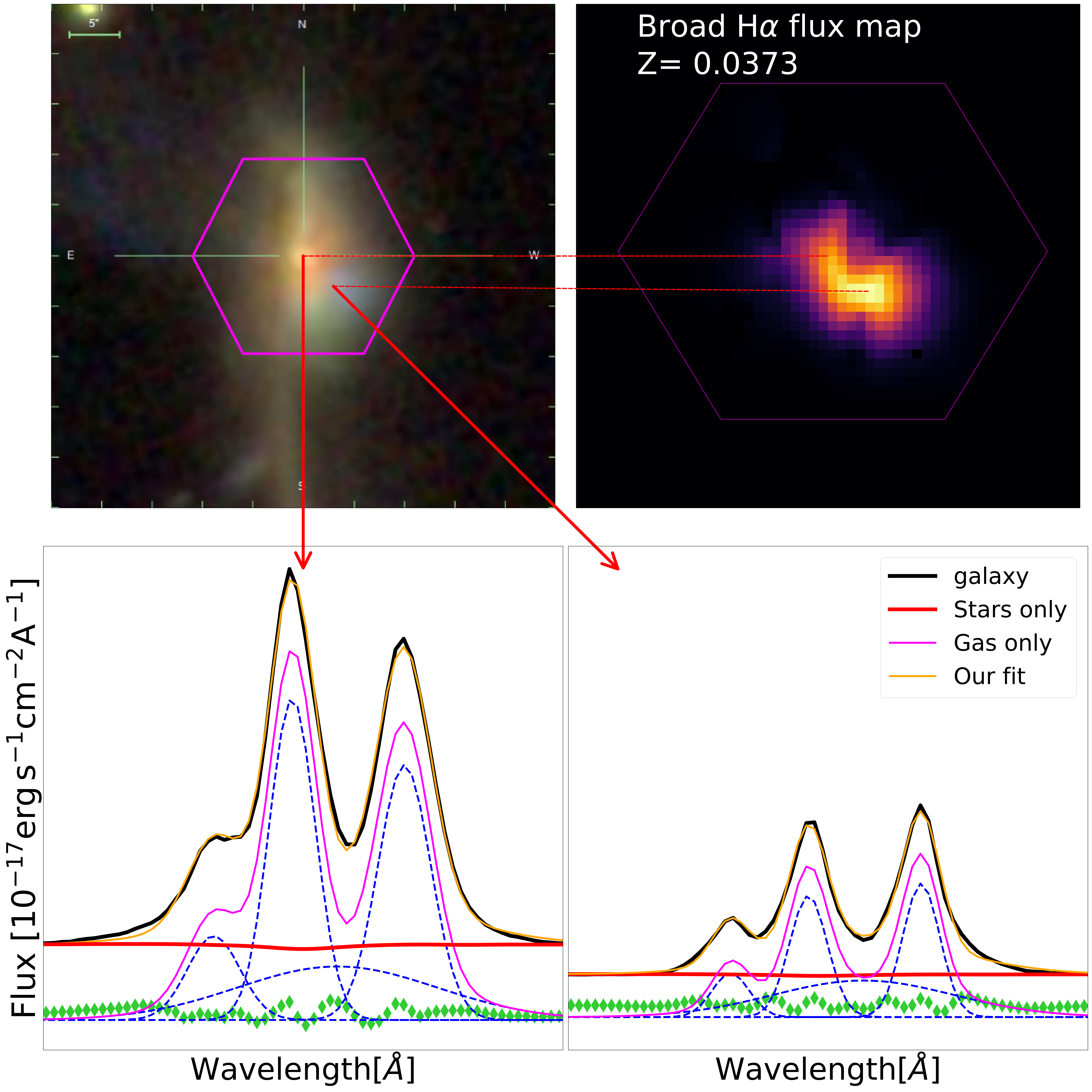}
    \caption{Image and emission line fitting result for dual AGN candidates (MaNGA-ID: 1-244377). The upper left panel is the galaxy image from MaNGA. The upper right panel is the map of broad $\rm H\alpha$ flux. The two galaxies are too close to be separated on the radio map. The two images below shows $\rm H\alpha$ fitting at the centres. }
    \label{fig:dualAGN11013}
\end{figure}

The currently largest catalogue of 36 dual AGNs (including 31 new findings and 5 previous cases) comes from \citet{dual_number_100,dual_number_50}. Such events are hard to identify and require high spatial resolution data to determine the presence of two AGNs in a galaxy rather than one large-scale AGN. We use a combination of broad $\rm H\alpha$ flux map and radio map from the VLA FIRST survey \citep{Firstsurvey} to confirm our detection of dual AGNs in MaNGA. In  \autoref{fig:dualAGN10218} and \autoref{fig:dualAGN8250}, we present two dual AGN mergers with clear bimodal radio features. In  \autoref{fig:dualAGN11013}, we present a dual AGN candidate where the two galaxies are too close to be separated on the radio map. But there is a clear bimodal feature on the broad $\rm H\alpha$ flux map.

Detection of such dual AGN events will help us to further understand the role of active galactic nuclei in galaxy evolution. In future work, it is possible to further investigate what kind of galactic nuclei will be excited as AGNs in mergers and how dual AGNs can form in the future.

\subsection{A combination of broad $\rm H\alpha$ and double narrow lines}
\label{sec:combination of lines}
\begin{figure*}
	\includegraphics[width=\textwidth]{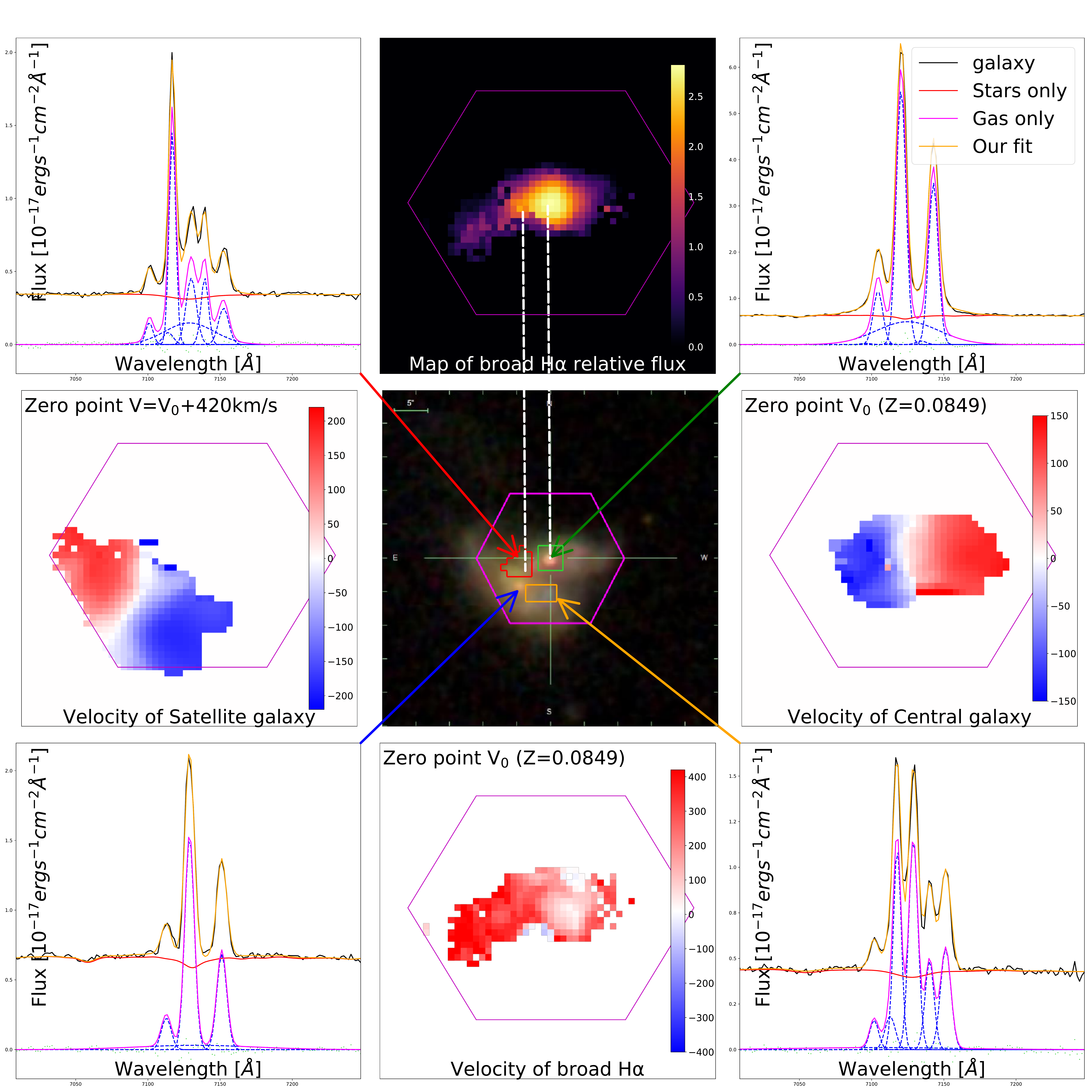}
    \caption{The case where broad H$\alpha$ and double narrow line features occur simultaneously. The MaNGA-ID of the object is 1-150947. The panels on the four corners represent the $\rm H\alpha$ spectra at different locations of the galaxy. The central box is its optical image. The panel on top of the optical image is the map of relative broad $\rm H\alpha$ flux and the panel at the bottom is the map of broad $\rm H\alpha$ velocity. The panels to the left and right are the narrow-line velocity maps of the central and satellite galaxy.}
    \label{fig:dualAGN10508}
\end{figure*}

In \autoref{sec:methods}, we use different templates to fit different multi-Gaussian features. We usually assume broad lines and double narrow lines are caused by different mechanisms. So we did not discuss the cases where they occur simultaneously. To the bottom right of  \autoref{fig:Hares} and \autoref{fig:OIIIres}, we still notice such feature in the spectrum of a merger sample shown in  \autoref{fig:dualAGN10508}.

According to the velocity maps of the central and satellite galaxy, the velocity at the satellite galaxy centre is 420 km/s larger than that of the central galaxy. The broad $\rm H\alpha$ velocity map shows that the broad $\rm H\alpha$ velocity in region where broad H$\alpha$ and double narrow line features occur simultaneously (red region on optical image) follows the velocity of the satellite galaxy. This suggests that the broad lines in the region and the central broad lines (green region) may come from different physical processes. One possible explanation is that in the overlapping area, there are three separated gas region along line of sight. One comes from the satellite galaxy and produces the narrow-line set with larger velocity. One comes from the central galaxy and produces the narrow-line set with lower velocity. And a region between them dominated by shock waves triggered by merging. But there is a possibility of a wandering black hole accreting in the overlapping region. This needs to be tested with higher resolution X-ray or radio images in future work. 

\section{Summary}
We have carried a systematic search for broad $\rm H\alpha$ and double-peaked narrow emission line features in the full MaNGA sample. Galaxies with $\rm H\alpha$ amplitude over noise ratio in larger than three are divided uniformly according to their mean g-band weighted signal-to-noise ratios. We identified 188 galaxies where single Gaussian emission line fits fail among 1$\rm \sigma$ outliers in each bin. There are 38 galaxies with broad $\rm H\alpha$ and [OIII] $\rm \lambda$5007 lines, 101 galaxies with broad $\rm H\alpha$ lines but no broad [OIII] $\rm \lambda$5007 lines, and 49 galaxies with double-peaked narrow emission lines. New emission line properties (including velocities, velocity dispersions, flux) are obtained by fitting multi-Gaussian templates in {\sc ppxf} \citep{Cappellari2022} with kinematic constraints on broad and narrow components. The galaxies with broad emission lines are classified into three catagories : central star-forming, central AGN, and diffuse central LINER according to their BPT diagram. 

The catalogue helps us further understand the AGN-galaxy coevolution through the stellar population of broad-line region host galaxies and connecting broad lines’ properties to the host galaxies’ dynamical properties. type-1 AGN occupy the full range of ages from old systems to young galaxies. The star-forming galaxies are relatively young and
LINER galaxies are usually old. Our work indicates that the host galaxy of type-1 AGNs are slightly younger than seyfert-2 galaxies while their metallicity distribution follows the same trend. This is consistent with results recently reported in \cite{AGN_age}.

We estimate the masses of black hole in type-1 AGN through broad $\rm H\alpha$ width and luminosity based on the formula given by \cite{M-broad}. Using the $\sigma_{\rm e}$ from DAP, we obtain our $M_{\rm BH}-\sigma_{\rm e}$ relation for type-1 AGN in \autoref{eq:MSrelation}. Finally, we discover three dual AGN candidates in mergers in \autoref{sec:dual AGN} and present one possible candidate for a wandering black hole in \autoref{sec:combination of lines}. This sample may be useful for further studies on AGN activities, feedback processes and their connections to stellar populations and dynamical properties.